\documentclass[12pt,letter]{article}
\pdfoutput=1
\usepackage{graphicx, epsfig, color,cite}
\usepackage{amsmath,amsthm,amsfonts,geometry,etoolbox}

\usepackage{amssymb}
\usepackage{caption,subcaption,graphicx,appendix}
\usepackage[hidelinks]{hyperref}
\def\to{\rightarrow}

\def\bi{\begin{itemize}}
\def\ei{\end{itemize}}
\def\te{\tilde e}

\def\tchi{\tilde\chi}

\def\tu{\tilde u}

\def\tb{\tilde b}

\def\tst{\tilde t}
\def\ttau{\tilde \tau}

\def\tg{\tilde g}
\def\tnu{\tilde\nu}

\def\tw{\widetilde\chi^{\pm}}

\def\tz{\widetilde\chi^0}
\def\alt{\lesssim}
\def\agt{\gtrsim}
\def\be{\begin{equation}}  
\def\ee{\end{equation}}  
\def\bea{\begin{eqnarray}}  
\def\eea{\end{eqnarray}}

\newcommand{\p}{\partial}
\newcommand{\myeq}{\begin{small}\begin{equation}\begin{aligned}}
\newcommand{\myeqend}{\end{aligned}\end{equation}\end{small}}

\begin{document}
\begin{titlepage}
\begin{flushright}
OU-HEP-211130
\end{flushright}

\vspace{0.5cm}
\begin{center}
{\Large \bf Comparison of SUSY spectra generators for \\
natural SUSY and string landscape predictions
}\\
\vspace{1.2cm} \renewcommand{\thefootnote}{\fnsymbol{footnote}}
{\large Howard Baer$^{1,2}$\footnote[1]{Email: baer@ou.edu },
Vernon Barger$^2$\footnote[2]{Email: barger@pheno.wisc.edu} and
Dakotah Martinez$^1$\footnote[4]{Email: dakotah.s.martinez-1@ou.edu} 
}\\ 
\vspace{1.2cm} \renewcommand{\thefootnote}{\arabic{footnote}}
{\it 
$^1$Homer L. Dodge Department of Physics and Astronomy,\\
University of Oklahoma, Norman, OK 73019, USA \\[3pt]
}
{\it 
$^2$Department of Physics,
University of Wisconsin, Madison, WI 53706 USA \\[3pt]
}

\end{center}

\vspace{0.5cm}
\begin{abstract}
\noindent
Models of natural supersymmetry give rise to a weak scale 
$m_{weak}\sim m_{W,Z,h}\sim 100$ GeV without any (implausible) finetuning
of independent contributions to the weak scale. These models, 
which exhibit radiatively driven naturalness (RNS), 
are expected to arise from statistical analysis of the string landscape 
wherein large soft terms are favored, but subject to a not-too-large 
value of the derived weak scale in each pocket universe of the greater multiverse.
The string landscape picture then predicts, using the Isajet SUSY spectra generator
Isasugra, a statistical peak at $m_h\sim 125$ GeV with sparticles generally
beyond current LHC search limits. In this paper, we investigate how well these
conclusions hold up using other popular spectra generators: SOFTSUSY, SPHENO and 
SUSPECT (SSS).
We built a computer code DEW4SLHA which operates on SUSY Les Houches Accord files to
calculate the associated electroweak naturalness measure $\Delta_{EW}$.
The SSS generators tend to yield a Higgs mass peak $\sim 125-127$ GeV with 
a superparticle mass spectra rather similar to that generated by Isasugra.
In an Appendix, we include loop corrections to $\Delta_{EW}$ in a more standard 
notation.
\end{abstract}
\end{titlepage}

\section{Introduction}
\label{sec:intro}

Supersymmetrization of the Standard Model (SM) elegantly solves the gauge hierarchy
problem (stabilizing the newly discovered Higgs boson mass under quantum corrections)
but at the expense of including a host of new matter states, the so-called superpartners.
Early expectations from naturalness predicted superpartners at or around the
weak scale\cite{Ellis:1986yg,Barbieri:1987fn,Dimopoulos:1995mi,Anderson:1994tr}. 
For instance, the naturalness upper bound for the gluino
was predicted (under the naturalness measure $\Delta_{BG}\alt 30$) to be
$m_{\tg}\alt 400$ GeV. In contrast, the current mass limits from LHC Run 2 searches with
139 fb$^{-1}$ claim $m_{\tg}\agt 2.25$ TeV\cite{ATLAS:2020syg,CMS:2019zmd}. 
The yawning gap between the weak scale
and the superpartner mass scale-- the Little Hierarchy Problem (LHP)\cite{Barbieri:2000gf}-- 
has lead many authors to conclude\cite{Lykken:2014bca,Dine:2015xga,Craig:2013cxa} that the weak scale supersymmetry\cite{Baer:2006rs} 
hypothesis is under intense pressure, and possibly even excluded.

However, it has been pointed out that the resolution to the LHP lies instead 
in that conventional early measures of naturalness over-estimated 
the finetuning\cite{Baer:2013gva,Mustafayev:2014lqa,Baer:2014ica}.
The BG log derivative measure\cite{Barbieri:1987fn}, $\Delta_{BG}\equiv max_i|\frac{\partial\log m_Z^2}{\partial\log p_i}|$ where the $p_i$ are fundamental parameters of the low energy effective field theory (EFT), depends strongly on what one assumes are independent parameters. To derive the bounds in Ref's \cite{Ellis:1986yg,Barbieri:1987fn,Dimopoulos:1995mi,Anderson:1994tr}, 
the authors adopted
common scalar masses $m_0$, gauginos masses $m_{1/2}$ and $A$-terms as 
independent parameters. However, in more ultraviolet complete theories, such as 
string theory, these parameters are all correlated. Adopting correlated soft terms
then greatly reduces the amount of finetuning which is calculated, often by 1-2 
orders of magnitude. An alternative measure $\Delta_{HS}\equiv \delta m_{H_u}^2/m_h^2$, 
(which is inconsistent with $\Delta_{BG}$ in that it splits $m_{H_u}^2(weak)$ into
$m_{H_u}^2(HS)+\delta m_{H_u}^2$ which destroys the focus point behavior inherent in 
$\Delta_{BG}$) discards RG contributions which show the interdependence of $m_{H_u}^2$
and $\delta m_{H_u}^2$.

An alternative measure for naturalness $\Delta_{EW}$ was proposed in \cite{Baer:2012up,Baer:2012cf} 
based on the notion of {\it practical naturalness}\cite{Baer:2015rja}: 
that all {\it independent} contributions to an observable ${\cal O}$ should be 
comparable to or less than ${\cal O}$. For 
instance, if ${\cal O}=o_1+\cdots +o_n$ where the $o_i$ are independent contributions to 
${\cal O}$, and if $o_1\gg {\cal O}$, then some other unrelated contribution
would have been a huge opposite sign contribution of precisely the right value such as to
maintain ${\cal O}$ at its measured value. Such finetunings, while logically possible, 
are thought to be {\it highly implausible} unless the contributions $o_i$ are related
by some symmetry, in which case they would not actually be independent. 
Practical naturalness has been successfully applied for instance by Gaillard and Lee
in the case of the $K_L-K_S$ mass difference to correctly predict the value of the
charm quark mass\cite{Gaillard:1974hs}. 
It is also closely related to {\it predictivity} in physical
theories in that missing contributions to an observable, such as higher order corrections in perturbation theory, should be (hopefully) small so that leading order terms 
provide a reliable estimate to any perturbatively calculated observable.

The minimization conditions for the MSSM Higgs potential allows one to relate the 
observed value of the weak scale to terms in the minimal 
supersymmetric standard model (MSSM) Lagrangian:
\be
m_Z^2/2=\frac{(m_{H_d}^2+\Sigma_d^d)-(m_{H_u}^2+\Sigma_u^u)\tan^2\beta}{\tan^2\beta -1}-\mu^2 
\label{eq:mzs}
\ee
where $m_{H_u}^2$ and $m_{H_d}^2$ are Higgs sector soft breaking masses, $\mu$ is the
(SUSY-conserving) $\mu$ parameter and $\tan\beta=v_u/v_d$ is the ratio of Higgs field
vacuum expectation values. The $\Sigma_d^d$ and $\Sigma_u^u$ terms contain a 
variety of loop corrections to the Higgs potential and are detailed in Ref. \cite{Baer:2012cf}
in the notation of {\it Weak Scale Supersymmetry} (WSS)\cite{Baer:2006rs} and given in the Appendix of this paper in the more
standard notation from S. P. Martin\cite{Martin:1997ns}. The most important of the loop
corrections typically comes from the top-squark sector, $\Sigma_u^u(\tst_{1,2})$.
Note that all contributions in Eq. \ref{eq:mzs} are evaluated at the weak scale
typically taken as $Q^2=m_{\tst_1}m_{\tst_2}$ such as to minimize the logs which are present in the $\Sigma_u^u(\tst_{1,2})$ contributions.

The $\Delta_{EW}$ measure is defined as
\be
\Delta_{EW}\equiv |{\rm largest\ contribution\ to\ RHS\ of\ Eq.~\ref{eq:mzs}}|/(m_Z^2/2).
\ee
One can quickly read off the consequences for a low value of $\Delta_{EW}$:
\bi 
\item $m_{H_u}^2$, which in the decoupling limit functions like the SM Higgs doublet 
and gives mass to the $W$, $Z$ and $h$ bosons, 
must be driven under radiative EWSB to {\it small} negative values, a condition known as
radiatively-driven naturalness (RNS). Thus, electroweak symmetry is {\it barely broken}.
\item The $\mu$ parameter, which feeds mass to the $W$, $Z$ and $h$ bosons as well as to the higgsinos, must be within a factor of several of $m_{W,Z,h}\sim 100$ GeV.
\item $m_A\sim m_{H_d}$ in the decoupling limit can live in the TeV regime since
the contribution of $m_{H_d}^2$ is suppressed by a factor $\tan^2\beta$.
\item Top squark contributions to the weak scale are loop suppressed
and so can live in the TeV range while maintaining naturalness.
\item The gluino contributes at two-loops\cite{Dedes:2002dy} and via RG running contributions to the
stop soft masses\cite{Brust:2011tb,Papucci:2011wy} and so also can live in the TeV range,
\item First and second generation sfermion contributions to the weak scale are via
Yukawa-suppressed 1-loop terms and via 2-loop RG contributions (which are dominant)\cite{Baer:2000xa}.
Thus, they can live in the 10-50 TeV regime which helps solve the SUSY flavor and CP problems\cite{Baer:2019zfl}.
\ei

An advantage of $\Delta_{EW}$ is that it is model independent insofar as it 
only depends on the weak scale sparticle and Higgs mass spectrum and not on how they are arrived at. Thus, a given spectrum will generate the same value of $\Delta_{EW}$ 
whether it was computed from the pMSSM or some high scale model. Also, requiring the contributions to $m_Z^2/2$ to be comparable to or less than its measured value
typically corresponds to an upper limit of $\Delta_{EW}\alt 30$. 
The turn-on of finetuning for $\Delta_{EW}\agt 30$ is visually displayed in 
Fig. 1 of Ref. \cite{Baer:2015rja}.

While WSS seems ruled out under the older naturalness measures\cite{Ellis:1986yg,Barbieri:1987fn,Dimopoulos:1995mi,Anderson:1994tr}, there is
still plenty of natural parameter space left unexplored by LHC under the $\Delta_{EW}$
measure\cite{Baer:2020kwz}. However, the $\Delta_{EW}$ measure does predict the existence of light higgsino-like EWinos $\tchi_1^\pm$ and $\tchi_{1,2}^0$ with mass $\sim 100-350$ GeV. 
The light higgsinos can be produced at decent rates at LHC, but owing to their small
mass gaps $m_{\tchi_2^0}-m_{\tchi_1^0}\sim 5-10$ GeV, there is only small visible energy
released in their decays, making detection a difficult\cite{Baer:2011ec} (but not impossible\cite{Baer:2014kya,Baer:2021srt})
prospect. The higgsino-like LSP $\tchi_1^0$ is thermally underproduced as dark matter,
leaving room for axionic dark matter as well\cite{Baer:2011hx}.

The $\Delta_{EW}$ naturalness measure is built in to the Isajet/Isasugra\cite{Paige:2003mg,Baer:1994nc}
event/spectrum generator. Also, the crucial 1-loop corrections to the Higgs potential
have been calculated within the (non-standard) notation of WSS\cite{Baer:2012cf}. 
As a result, of the spectrum generators available, 
Isasugra has been used the most for such studies. These include sparticle mass
bounds from naturalness, and parameter space limits and lucrative collider 
signatures from natural SUSY. However, a variety of other SUSY/Higgs spectra 
generators are available, including SUSPECT\cite{Djouadi:2002ze}, SOFTSUSY\cite{Allanach:2001kg} and SPHENO\cite{Porod:2003um}.
Some special Higgs spectrum calculators include FeynHIGGS\cite{Bahl:2018qog} and 
SUSYHD\cite{PardoVega:2015eno} and others\cite{Slavich:2020zjv}. 
Thus, it would be useful to know how other spectrum generators
compare to Isasugra in their natural SUSY spectra. For this reason, we have 
built a computer code DEW4SLHA which operates on a SUSY Les Houches Accord file 
(SLHA)\cite{Skands:2003cj} which is the standard output of spectrum generators. The program
computes the associated value of $\Delta_{EW}$ and all the various contributions.
In Sec. \ref{sec:code} of this paper, we introduce the code DEW4SLHA along with pointers on its accessibility.

While natural SUSY is highly interesting in its own right, some authors maintain 
that naturalness should cede ground to the emergent landscape/multiverse 
picture of string theory: if the cosmological constant $\Lambda_{cc}$ is finetuned 
to tiny values via anthropic selection in the multiverse, then why not also the 
weak scale? There is expected to be a statistical pull to large soft SUSY breaking terms
via a power law\cite{Douglas:2004qg,Susskind:2004uv,Arkani-Hamed:2005zuc} or log distribution\cite{Broeckel:2020fdz} in the landscape of string theory vacua.
However, one of the most important predictions of SUSY theories is the magnitude of the weak scale $m_{weak}$. Agrawal {\it et al.}\cite{Agrawal:1997gf,Agrawal:1998xa} have shown that if the 
pocket universe value of the weak scale is greater than a factor of 2-5 times our
universe's measured value, then complex nuclei, and hence atoms as we know them, would not arise. Now in a subset of vacua with the MSSM as low energy EFT but with
variable soft terms, then absent finetuning, the pocket universe value of the 
weak scale $m_{weak}^{PU}$ will nearly be the maximal contribution to the RHS of
Eq. \ref{eq:mzs}. Thus, a value $m_Z^{PU}\sim 4m_Z^{OU}$ corresponds to a 
value $\Delta_{EW}\alt 30$. This {\it anthropic} veto has been used along with a 
landscape pull to large soft terms to make statistical predictions from the
 string landscape for the SUSY and Higgs boson masses. It is found using Isasugra 
that the Higgs mass $m_h$ rises to a peak at $m_h\sim 125$ GeV while sparticles
such as the lightest stop and gluino are pulled to values beyond LHC13 search limits.
It would also be of interest to confirm or refute these results using other 
spectra/Higgs mass calculators.

Thus, in this paper we first introduce the public code DEW4SLHA in Sec. \ref{sec:code}.
In Sec. \ref{sec:BMpoints}, we apply this code to a natural SUSY benchmark point to
compare spectra from Isasugra against results from SOFTSUSY, SUSPECT and SPHENO.
As such, our paper follows previous comparison work but within the 
context of natural SUSY and string landscape phenomenology\cite{Allanach:2002pz,Belanger:2005jk}.
In Sec. \ref{sec:BMpoints}, we also move beyond benchmark points to compare Higgs mass and naturalness contours in the scalar mass vs. gaugino mass parameter planes for
just the SOFTSUSY spectrum generator.
In Sec. \ref{sec:landscape}, we use SOFTSUSY to generate statistical landscape
predictions to compare against earlier work from Isasugra.
A summary and conclusions are given in Sec. \ref{sec:conclude}. 
In an Appendix \ref{sec:appendix}, we present expressions for the $\Sigma_u^u$ and $\Sigma_d^d$ 
contributions in the standard notation of S. P. Martin's 
SUSY primer \cite{Martin:1997ns}.
 
\section{The DEW4SLHA code}
\label{sec:code}

A new code, DEW4SLHA, has been developed in Python 3 by D. Martinez to evaluate $\Delta_{EW}$ from any user-supplied SLHA-format output from a spectrum generator such as Isajet, SOFTSUSY, SUSPECT, or SPHENO. The standalone executable can be found at 
\verb|https://dew4slha.com|, along with instructions on how to run the program from a Linux terminal. The source code can be found at 
\verb| https://github.com/Dmartinez-96/DEW-Calculator |. 
The DEW4SLHA code uses the SLHA particle/sparticle pole masses from block MASS and the
running soft term values from block MSOFT.
DEW4SLHA has the capability to operate on SLHA files with a single-scale
output or a grid of outputs, with the number of grid points specified by
Switch 11 in the SLHA block MODSEL.
In the case of the latter, DEW4SLHA extracts the values of parameters
at the maximum grid scale and computes DEW using the parameters at this scale. 
The computational routine of the program follows the equations presented in the Appendix \ref{sec:appendix} and then orders the 44 1-loop contributions to the Higgs minimization condition by magnitude. 
Two corrections at the 2-loop level are included in the routine to include the effects of the gluino mass on the DEW measure\cite{Dedes:2002dy}. Similar codes have been developed but are not to our knowledge publicly available\cite{vanBeekveld:2016hug,vanBeekveld:2019tqp}.

\section{Natural SUSY benchmark points}
\label{sec:BMpoints}

\begin{figure}[!htbp]
\begin{center}
\includegraphics[height=0.4\textheight]{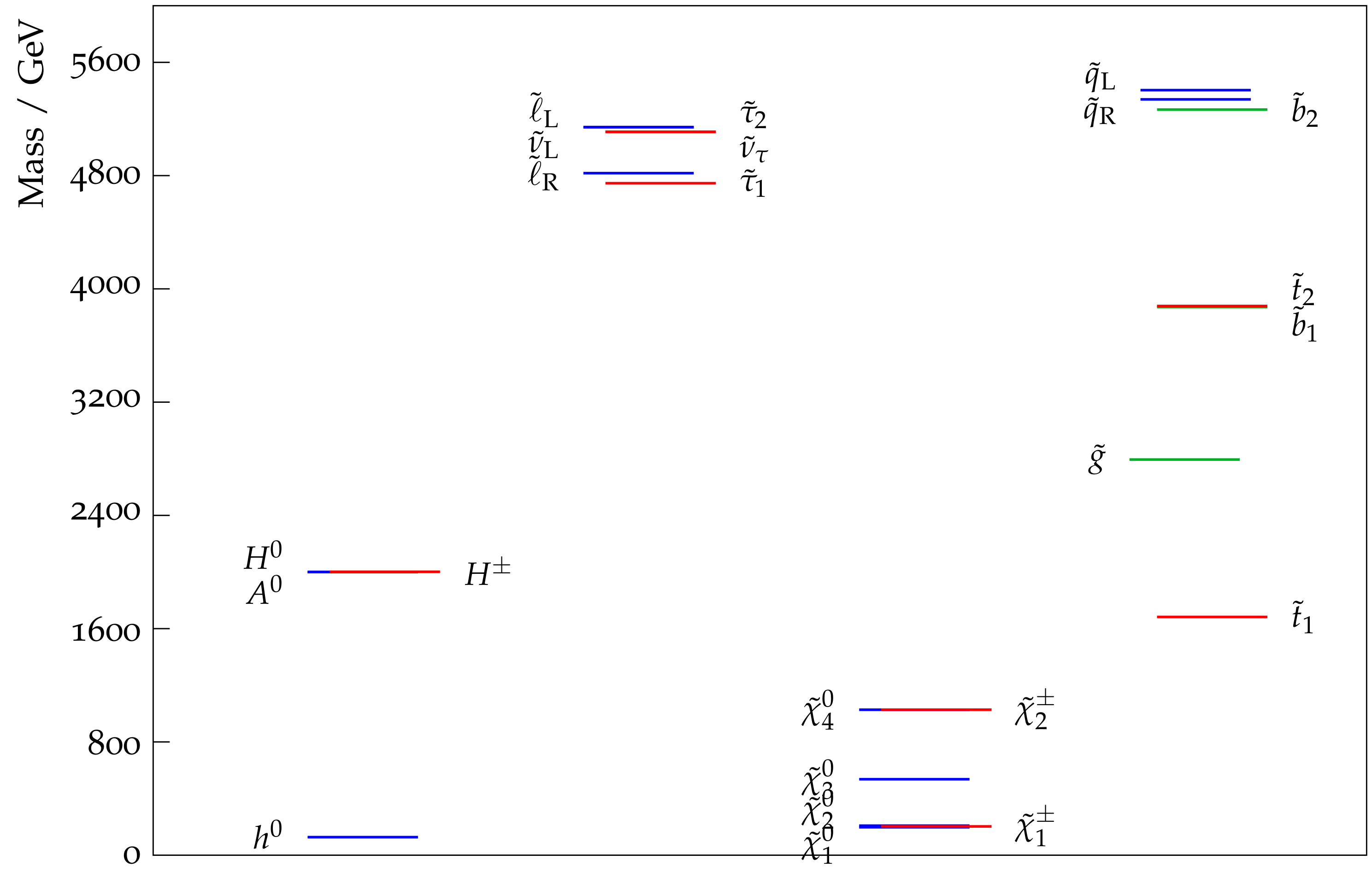}
\caption{Sparticle and Higgs mass spectra for a natural SUSY
benchmark point from SOFTSUSY.
\label{fig:bm}}
\end{center}
\end{figure}

Using the code DEW4SLHA, we can now compare spectra generated from the 
various spectra calculators for a particular natural SUSY benchmark point.
For the BM point, we adopt the two-extra-parameter non-universal Higgs model (NUHM2)\cite{Ellis:2002iu,Baer:2005bu}
with input parameters
\be
m_0,\ m_{1/2},\ A_0,\ \tan\beta,\ \mu,\ m_A
\ee
where we have traded the high scale Higgs soft masses $m_{H_u}^2$ and $m_{H_d}^2$ 
for the more convenient weak scale parameters $\mu$ and $m_A$.
Then we adopt the benchmark parameter values $m_0=5$ TeV, $m_{1/2}=1.2$ TeV, 
$A_0=-8$ TeV, $\tan\beta =10$, $\mu =200$ GeV and $m_A=2$ TeV.
A pictorial representation of the spectra using SOFTSUSY is shown in 
Fig.~\ref{fig:bm} where we see that indeed the higgsinos and Higgs boson $h$
lie in the $100-200$ GeV range whilst the top-squarks and gluino live in the
several TeV regime.

In Table \ref{tab:bm}, we list the mass spectra and $\Delta_{EW}$ values from 
each of four spectra generators.
For ISAJET, we use version 7.88\cite{Paige:2003mg} while for SUSPECT we use version 2.51\cite{Djouadi:2002ze}.
For SOFTSUSY, we use version 4.1.10\cite{Allanach:2001kg} including two-loop corrections
to $m_{\tg}$ and the default two-loop corrections to $m_h$.
We use SPHENO version 4.0.4\cite{Porod:2003um} with MSSM-to-SM matching at scale $Q=m_{SUSY}=\sqrt{m_{\tst_1}m_{\tst_2}}$. In contrast, SOFTSUSY imposes EFT matching at $Q=m_Z$ while ISAJET uses multiple scales\cite{Baer:2005pv}. 
The gluino masses are all within 1.5\% of each other.
The naturalness parameters for three codes are all less than thirty; the outlier here is SPHENO where also the light top squark mass $m_{\tst_1}$ is somewhat higher than 
the other codes. Here, the top squark masses are highly sensitive to mixing 
which comes from the weak scale value of $A_t$ and indeed the values of
$A_t(Q)$ for Isasugra/SOFTSUSY/SUSPECT/SPHENO are 
-4898/-4830/-4894/-5090 GeV, respectively. 
Thus, SPHENO has slightly more stop mixing than the other codes which increases $\Delta_{EW}$ somewhat. Another difference comes from the value of $m_h$ generated:
both SOFTSUSY and SUSPECT generate $m_h\sim 127.4$ GeV whilst SPHENO 
generates $m_h=125.2$ GeV and Isasugra generates $m_h=124.7$ GeV. It can be remarked 
that Isasugra has the least sophisticated light Higgs mass calculation, 
and includes only third generation sparticle 1-loop contributions to $m_h$.
Another feature is that the Isasugra value of $m_{\tchi_1^\pm}$ is about six 
GeV higher than SOFTSUSY and SUSPECT while the SPHENO is six GeV lower. 
These values depend sensitively on the scale choice at which each EWino mass 
is calculated. For instance, Isasugra uses the Pierce {\it et al.} (PBMZ)\cite{Pierce:1996zz}
recipe to calculate each mass separately at each mass scale.
\begin{table}\centering
\begin{tabular}{lcccc}
\hline
parameter & Isasugra & SOFTSUSY & SUSPECT & SPHENO \\
\hline
$m_{\tg}$   & 2830.7  & 2794.3 & 2838.6 & 2827.6 \\
$m_{\tu_L}$ & 5440.3 & 5403.2 & 5406.0 & 5412.8 \\
$m_{\tu_R}$ & 5561.7 & 5521.3 & 5523.0 & 5521.8 \\
$m_{\te_R}$ & 4823.0  & 4817.3 & 4818.1 & 4825.8 \\
$m_{\tst_1}$ & 1714.3 & 1682.8 & 1746.9 & 1942.1 \\
$m_{\tst_2}$ & 3915.1 & 3879.0 & 3899.2 & 3947.0 \\
$m_{\tb_1}$ & 3949.1 & 3871.6 & 3891.7 & 3939.1 \\
$m_{\tb_2}$ & 5287.5 & 5266.4 & 5277.2 & 5281.7 \\
$m_{\ttau_1}$ & 4745.7 & 4746.1 & 4749.1 & 4757.4 \\
$m_{\ttau_2}$ & 5110.2 & 5109.7 & 5110.8 & 5107.2 \\
$m_{\tnu_{\tau}}$ & 5116.8 & 5108.7 & 5113.8 & 5106.2 \\
$m_{\tw_2}$ & 1020.2 & 1027.5 & 1030.6 & 1031.9 \\
$m_{\tw_1}$ & 209.7 & 203.1 & 203.0 & 197.3 \\
$m_{\tz_4}$ & 1033.5 & 1027.3 & 1031.1 & 1032.0 \\ 
$m_{\tz_3}$ & 540.1 & 536.4 & 537.2 & 538.1 \\ 
$m_{\tz_2}$ & -208.3 & -208.6 & -208.7 & -203.0 \\ 
$m_{\tz_1}$ & 197.9 & 197.2 & 197.1 & 191.9 \\ 
$m_h$       & 124.7 & 127.3 & 127.5 & 125.2 \\ 
\hline
$\Delta_{\rm EW}$ & 24.8 & 23.0 & 28.2  & 44.1 \\
\hline
\end{tabular}
\caption{Sparticle and Higgs mass spectra from four spectra generators for a natural SUSY benchmark point with $m_0=5$ TeV, $m_{1/2}=1.2$ TeV, $A_0=-8$ TeV, $\tan\beta =10$
with $\mu =200$ GeV and $m_A=2$ TeV.
}
\label{tab:bm}
\end{table}

In Table \ref{tab:Sigmas}, we list the top 46 contributions to $\Delta_{EW}$ 
from each of the spectra codes. We see from line 1 that the largest contribution
comes for each code from $\Sigma_u^u(\tst_2)$ which sets the value of $\Delta_{EW}$,
and where we see that SPHENO gives the largest value. The second largest
contribution comes from $\Sigma_u^u(\tst_1)$ as might be expected.
The next several largest contributions come from $H_d$, $\mu$ and $H_u$
and $\Sigma_u^u(\tb_{1,2})$ although the ordering of these differs among the codes.
In general, the agreement for the remaining contributions is typically 
within expectations.
\newpage
%
%
%
\begin{table}[h!]
    \begin{center}
		\resizebox{0.85\hsize}{!}{\begin{tabular}{c c c c c} 
			\hline
			Order & Isajet & SoftSUSY & Suspect  & Spheno\\ [0.5ex] 
			\hline
			1 & 24.819, $\Sigma_{u}^{u}(\widetilde{t}_{2})$ & 23.015, $\Sigma_{u}^{u}(\widetilde{t}_{2})$ & 28.227, $\Sigma_{u}^{u}(\widetilde{t}_{2})$ & 44.062, $\Sigma_{u}^{u}(\widetilde{t}_{2})$\\
			2 & 19.367, $\Sigma_{u}^{u}(\widetilde{t}_{1})$ & 18.318, $\Sigma_{u}^{u}(\widetilde{t}_{1})$ & 20.372, $\Sigma_{u}^{u}(\widetilde{t}_{1})$ & 27.465, $\Sigma_{u}^{u}(\widetilde{t}_{1})$\\
			3 & 10.449, $\Sigma_{u}^{u}(\mathcal{O}(\alpha_{s}\alpha_{t}))$ & 10.074, $H_{d}$ & 10.294, $H_{d}$ & 11.205, $H_{u}$\\
			4 & 10.424, $H_{d}$ & 9.618, $\mu$ & 9.621, $\mu$ & 10.298, $H_{d}$\\
			5 & 9.625, $\mu$ & 6.985, $\Sigma_{u}^{u}(\mathcal{O}(\alpha_{s}\alpha_{t}))$ & 7.405, $\Sigma_{u}^{u}(\mathcal{O}(\alpha_{s}\alpha_{t}))$ & 9.621, $\mu$\\
			6 & 5.861, $H_{u}$ & 4.557, $\Sigma_{u}^{u}(\widetilde{b}_{2})$ & 4.044, $\Sigma_{u}^{u}(\widetilde{b}_{2})$ & 8.321, $\Sigma_{u}^{u}(\mathcal{O}(\alpha_{s}\alpha_{t}))$\\
			7 & 4.164, $\Sigma_{u}^{u}(\widetilde{\tau}_{2})$ & 4.316, $\Sigma_{u}^{u}(\widetilde{\tau}_{2})$ & 3.761, $\Sigma_{u}^{u}(\widetilde{\tau}_{2})$ &  3.604, $\Sigma_{u}^{u}(\widetilde{b}_{1})$\\
			8 & 3.933, $\Sigma_{u}^{u}(\widetilde{b}_{2})$ & 3.252, $\Sigma_{u}^{u}(\widetilde{\tau}_{1})$ & 2.801, $\Sigma_{u}^{u}(\Sigma$ 2nd gen. $\widetilde{q})$ & 2.505, $\Sigma_{u}^{u}(\widetilde{\tau}_{2})$\\
			9 & 2.970, $\Sigma_{u}^{u}(\widetilde{\tau}_{1})$ & 2.909, $\Sigma_{u}^{u}(\Sigma$ 2nd gen. $\widetilde{q})$ & 2.801, $\Sigma_{u}^{u}(\Sigma$ 1st gen. $\widetilde{q})$ & 2.486, $\Sigma_{u}^{u}(\widetilde{b}_{2})$\\
			10 & 2.912, $\Sigma_{u}^{u}(\Sigma$ 2nd gen. $\widetilde{q})$ & 2.909, $\Sigma_{u}^{u}(\Sigma$ 1st gen. $\widetilde{q})$ & 2.653, $\Sigma_{u}^{u}(\widetilde{\tau}_{1})$ & 2.468, $\Sigma_{u}^{u}(\Sigma$ 2nd gen. $\widetilde{q})$\\
			11  & 2.912, $\Sigma_{u}^{u}(\Sigma$ 1st gen. $\widetilde{q})$ & 2.761, $H_{u}$ & 2.507, $\Sigma_{u}^{u}(\widetilde{b}_{1})$ & 2.468, $\Sigma_{u}^{u}(\Sigma$ 1st gen. $\widetilde{q})$\\
			12 & 2.003, $\Sigma_{u}^{u}(\widetilde{b}_{1})$ & 2.101, $\Sigma_{u}^{u}(\widetilde{b}_{1})$ & 1.212, $\Sigma_{u}^{u}(\widetilde{\chi}_{2}^{\pm})$ & 1.263, $\Sigma_{u}^{u}(\widetilde{\chi}_{2}^{\pm})$\\
			13 & 1.169, $\Sigma_{u}^{u}(\widetilde{\chi}_{2}^{\pm})$ & 1.191, $\Sigma_{u}^{u}(\widetilde{\chi}_{2}^{\pm})$ & 9.235e-1, $\Sigma_{u}^{u}(\widetilde{Z}_{3}^{0})$ & 1.133, $\Sigma_{u}^{u}(\widetilde{\tau}_{1})$\\
			14 & 9.765e-1, $\Sigma_{u}^{u}(\widetilde{Z}_{3}^{0})$ & 9.114e-1, $\Sigma_{u}^{u}(\widetilde{Z}_{3}^{0})$ & 7.312e-1, $H_{u}$ & 9.538e-1, $\Sigma_{u}^{u}(\widetilde{Z}_{3}^{0})$\\
			15 & 6.987e-1, $\Sigma_{u}^{u}(\widetilde{Z}_{4}^{0})$ & 6.924e-1, $\Sigma_{u}^{u}(\widetilde{Z}_{4}^{0})$ & 7.076e-1, $\Sigma_{u}^{u}(\widetilde{Z}_{4}^{0})$ & 7.381e-1, $\Sigma_{u}^{u}(\widetilde{Z}_{4}^{0})$\\
			16 & 5.98e-1, $\Sigma_{u}^{u}(H^{\pm})$ & 6.083e-1, $\Sigma_{u}^{u}(H^{\pm})$ & 6.264e-1, $\Sigma_{u}^{u}(H^{\pm})$ & 6.755e-1, $\Sigma_{u}^{u}(H^{\pm})$\\
			17 & 1.532e-1, $\Sigma_{u}^{u}(t)$ & 1.438e-1, $\Sigma_{u}^{u}(t)$ & 1.440e-1, $\Sigma_{u}^{u}(t)$ & 2.064e-1, $\Sigma_{u}^{u}(\widetilde{Z}_{1}^{0})$\\
			18 & 5.924e-2, $\Sigma_{u}^{u}(\widetilde{Z}_{1}^{0})$ & 7.522e-2, $\Sigma_{u}^{u}(\widetilde{Z}_{1}^{0})$ & 7.687e-2, $\Sigma_{u}^{u}(\widetilde{Z}_{1}^{0})$ & 1.361e-1, $\Sigma_{u}^{u}(t)$\\
			19 & 5.543e-2, $\Sigma_{d}^{d}(H^{0})$ & 5.305e-2, $\Sigma_{d}^{d}(H^{0})$ & 5.564e-2, $\Sigma_{d}^{d}(H^{0})$ & 5.831e-2, $\Sigma_{d}^{d}(H^{0})$\\
			20 & 4.758e-2, $\Sigma_{d}^{d}(\widetilde{Z}_{3}^{0})$ & 4.397e-2, $\Sigma_{d}^{d}(\widetilde{Z}_{3}^{0})$ & 4.507, $\Sigma_{d}^{d}(\widetilde{Z}_{3}^{0})$ & 4.649e-2, $\Sigma_{d}^{d}(\widetilde{Z}_{3}^{0})$\\
			21 & 4.3e-2, $\Sigma_{u}^{u}(Z^{0})$ & 4.175e-2, $\Sigma_{d}^{d}(\widetilde{b}_{2})$ & 3.909e-2, $\Sigma_{d}^{d}(\widetilde{b}_{2})$ & 4.341e-2, $\Sigma_{d}^{d}(\widetilde{t}_{1})$\\
			22 & 4.3e-2, $\Sigma_{d}^{d}(\widetilde{b}_{2})$ & 3.783, $\Sigma_{u}^{u}(Z^{0})$ & 3.825e-2, $\Sigma_{u}^{u}(Z^{0})$ & 3.889e-2, $\Sigma_{u}^{u}(Z^{0})$\\
			23 & 3.748e-2, $\Sigma_{d}^{d}(\widetilde{t}_{1})$ & 3.438, $\Sigma_{d}^{d}(\widetilde{t}_{1})$ & 3.713e-2, $\Sigma_{d}^{d}(\widetilde{t}_{1})$ & 2.793e-2, $\Sigma_{d}^{d}(\widetilde{b}_{2})$\\
			24 & 3.198e-2, $\Sigma_{d}^{d}(\Sigma$ 2nd gen. $\widetilde{q})$ & 3.128e-2, $\Sigma_{d}^{d}(\Sigma$ 2nd gen. $\widetilde{q})$ & 3.075e-2, $\Sigma_{d}^{d}(\Sigma$ 2nd gen. $\widetilde{q})$ & 2.706e-2, $\Sigma_{d}^{d}(\Sigma$ 2nd gen. $\widetilde{q})$\\
			25 & 3.198e-2, $\Sigma_{d}^{d}(\Sigma$ 1st gen. $\widetilde{q})$ & 3.128e-2, $\Sigma_{d}^{d}(\Sigma$ 1st gen. $\widetilde{q})$ & 3.075e-2, $\Sigma_{d}^{d}(\Sigma$ 1st gen. $\widetilde{q})$ & 2.706e-2, $\Sigma_{d}^{d}(\Sigma$ 1st gen. $\widetilde{q})$\\
			26 & 2.329e-2, $\Sigma_{u}^{u}(h^{0})$ & 2.377e-2, $\Sigma_{u}^{u}(h^{0})$ & 2.395e-2, $\Sigma_{u}^{u}(h^{0})$ & 2.323e-2, $\Sigma_{u}^{u}(h^{0})$\\
			27 & 1.875e-2, $\Sigma_{d}^{d}(\widetilde{Z}^{0}_{4})$ & 1.841e-2, $\Sigma_{d}^{d}(\widetilde{Z}^{0}_{4})$ & 1.895e-2, $\Sigma_{d}^{d}(\widetilde{Z}^{0}_{4})$ & 2.152e-2, $\Sigma_{u}^{u}(\widetilde{Z}^{0}_{2})$\\
			28 & 1.669e-2, $\Sigma_{d}^{d}(\widetilde{\tau}_{1})$ & 1.787e-2, $\Sigma_{d}^{d}(\widetilde{\tau}_{1})$ & 1.504e-2, $\Sigma_{d}^{d}(\widetilde{\tau}_{1})$ & 1.974e-2, $\Sigma_{d}^{d}(\widetilde{Z}^{0}_{4})$\\
			29 & 1.279e-2, $\Sigma_{d}^{d}(\widetilde{\chi}_{2}^{\pm})$ & 1.276e-2, $\Sigma_{d}^{d}(\widetilde{\chi}_{2}^{\pm})$ & 1.326e-2, $\Sigma_{d}^{d}(\widetilde{\chi}_{2}^{\pm})$ & 1.719e-2, $\Sigma_{d}^{d}(\widetilde{Z}^{0}_{1})$\\
			30 & 1.102e-2, $\Sigma_{d}^{d}(\widetilde{\tau}_{2})$ & 1.107e-2, $\Sigma_{u}^{u}(H^{0})$ & 1.079e-2, $\Sigma_{u}^{u}(H^{0})$ & 1.553e-2, $\Sigma_{d}^{d}(\widetilde{b}_{1})$\\
			31 & 1.095e-2, $\Sigma_{d}^{d}(\mathcal{O}(\alpha_{s}\alpha_{t}))$ & 1.101e-2, $\Sigma_{d}^{d}(\widetilde{\tau}_{2})$ & 1.034e-2, $\Sigma_{d}^{d}(\widetilde{b}_{1})$ & 1.380e-2, $\Sigma_{d}^{d}(\widetilde{\chi}_{2}^{\pm})$\\
			32 & 9.869e-3, $\Sigma_{u}^{u}(\widetilde{Z}_{2}^{0})$ & 8.412e-3, $\Sigma_{d}^{d}(\widetilde{b}_{1})$ & 9.897e-3, $\Sigma_{d}^{d}(\widetilde{\tau}_{2})$ & 8.754e-3, $\Sigma_{u}^{u}(H^{0})$\\
			33 & 8.366e-3, $\Sigma_{d}^{d}(\widetilde{b}_{1})$ & 7.381e-3, $\Sigma_{d}^{d}(\mathcal{O}(\alpha_{s}\alpha_{t}))$ & 7.391e-3, $\Sigma_{d}^{d}(\mathcal{O}(\alpha_{s}\alpha_{t}))$ & 8.132e-3, $\Sigma_{d}^{d}(\mathcal{O}(\alpha_{s}\alpha_{t}))$\\
			34 & 8.083e-3, $\Sigma_{u}^{u}(H^{0})$ & 7.315e-3, $\Sigma_{u}^{u}(\widetilde{Z}_{2}^{0})$ & 7.180e-3, $\Sigma_{u}^{u}(\widetilde{Z}_{2}^{0})$ & 7.408e-3, $\Sigma_{d,u}^{d,u}(H^{\pm})$\\
			35 & 6.658e-3, $\Sigma_{d,u}^{d,u}(H^{\pm})$ & 6.542e-3, $\Sigma_{d,u}^{d,u}(H^{\pm})$ & 6.877e-3, $\Sigma_{d,u}^{d,u}(H^{\pm})$ & 6.470e-3, $\Sigma_{d}^{d}(\widetilde{\tau}_{2})$\\
			36 & 5.469e-3, $\Sigma_{u}^{u}(W^{\pm})$ & 5.400e-3, $\Sigma_{u}^{u}(W^{\pm})$ & 5.467e-3, $\Sigma_{u}^{u}(W^{\pm})$ & 6.324e-3, $\Sigma_{d}^{d}(\widetilde{\tau}_{1})$\\
			37 & 2.611e-3, $\Sigma_{d}^{d}(\widetilde{Z}_{2}^{0})$ & 2.660e-3, $\Sigma_{d}^{d}(\widetilde{Z}_{2}^{0})$ & 2.717e-3, $\Sigma_{d}^{d}(\widetilde{Z}_{2}^{0})$ & 5.561e-3, $\Sigma_{u}^{u}(W^{\pm})$\\
			38 & 1.081e-3, $\Sigma_{d}^{d}(\widetilde{Z}_{1}^{0})$ & 2.305e-3, $\Sigma_{d}^{d}(\widetilde{Z}_{1}^{0})$ & 2.428e-3, $\Sigma_{d}^{d}(\widetilde{Z}_{1}^{0})$ & 2.441e-3, $\Sigma_{u}^{u}(\widetilde{\chi}_{1}^{\pm})$\\
			39 & 7.420e-4, $\Sigma_{d}^{d}(h^{0})$ & 2.044e-3, $\Sigma_{u}^{u}(\widetilde{\chi}_{1}^{\pm})$ & 2.336, $\Sigma_{u}^{u}(\widetilde{\chi}_{1}^{\pm})$ & 1.630e-3, $\Sigma_{d}^{d}(\widetilde{Z}_{2}^{0})$\\
			40 & 4.723e-4, $\Sigma_{d,u}^{d,u}(Z^{0})$ & 7.568e-4, $\Sigma_{d}^{d}(h^{0})$ & 7.776e-4, $\Sigma_{d}^{d}(h^{0})$ & 7.394e-4, $\Sigma_{d}^{d}(h^{0})$\\
			41 & 4.205e-4, $\Sigma_{u}^{u}(\widetilde{\chi}_{1}^{\pm})$ & 4.069e-4, $\Sigma_{d,u}^{d,u}(Z^{0})$ & 4.199e-4, $\Sigma_{d,u}^{d,u}(Z^{0})$ & 4.265e-4, $\Sigma_{d,u}^{d,u}(Z^{0})$\\
			42 & 1.000e-4, $\Sigma_{d}^{d}(\widetilde{t}_{2})$ & 2.013e-4, $\Sigma_{d}^{d}(\widetilde{t}_{2})$ & 2.673e-4, $\Sigma_{d}^{d}(\widetilde{t}_{2})$ & 4.215e-4, $\Sigma_{d}^{d}(\widetilde{t}_{2})$\\
			43 & 6.007e-5, $\Sigma_{d,u}^{d,u}(W^{\pm})$ & 5.808e-5, $\Sigma_{d,u}^{d,u}(W^{\pm})$ & 6.002e-5, $\Sigma_{d,u}^{d,u}(W^{\pm})$ & 6.098e-5, $\Sigma_{d,u}^{d,u}(W^{\pm})$\\
			44 & 9.197e-6, $\Sigma_{d}^{d}(\widetilde{\chi}_{1}^{\pm})$ & 2.608e-5, $\Sigma_{d}^{d}(\widetilde{\chi}_{1}^{\pm})$ & 2.986e-5, $\Sigma_{d}^{d}(\widetilde{\chi}_{1}^{\pm})$ & 3.085e-5, $\Sigma_{d}^{d}(\widetilde{\chi}_{1}^{\pm})$\\
			45 & 2.315e-8, $\Sigma_{d}^{d}(b)$ & 2.302e-8, $\Sigma_{d}^{d}(b)$ & 2.282e-8, $\Sigma_{d}^{d}(b)$ & 1.895e-8, $\Sigma_{d}^{d}(b)$\\
			46 & 9.579e-9, $\Sigma_{d}^{d}(\tau)$ & 7.904e-9, $\Sigma_{d}^{d}(\tau)$ & 7.812e-9, $\Sigma_{d}^{d}(\tau)$ & 7.783e-9, $\Sigma_{d}^{d}(\tau)$\\
			\hline
		\end{tabular}}
	\end{center}
\caption{Top 46 contributions to $\Delta_{EW}$ for our natural SUSY benchmark point for four different spectra calculator codes.}
\label{tab:Sigmas}
\end{table}

\begin{figure}[!htbp]
\begin{center}
\includegraphics[height=0.28\textheight]{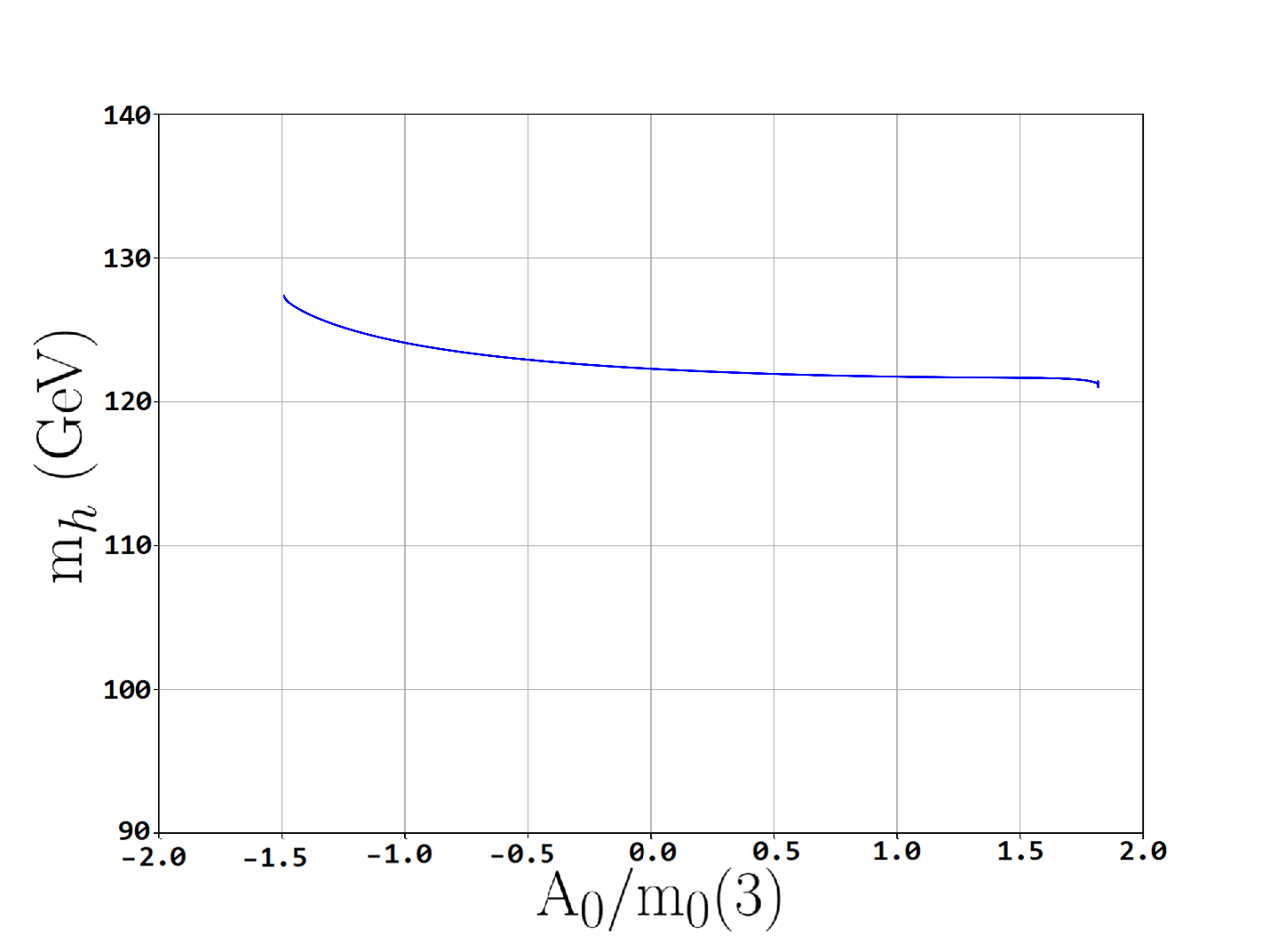}
\includegraphics[height=0.28\textheight]{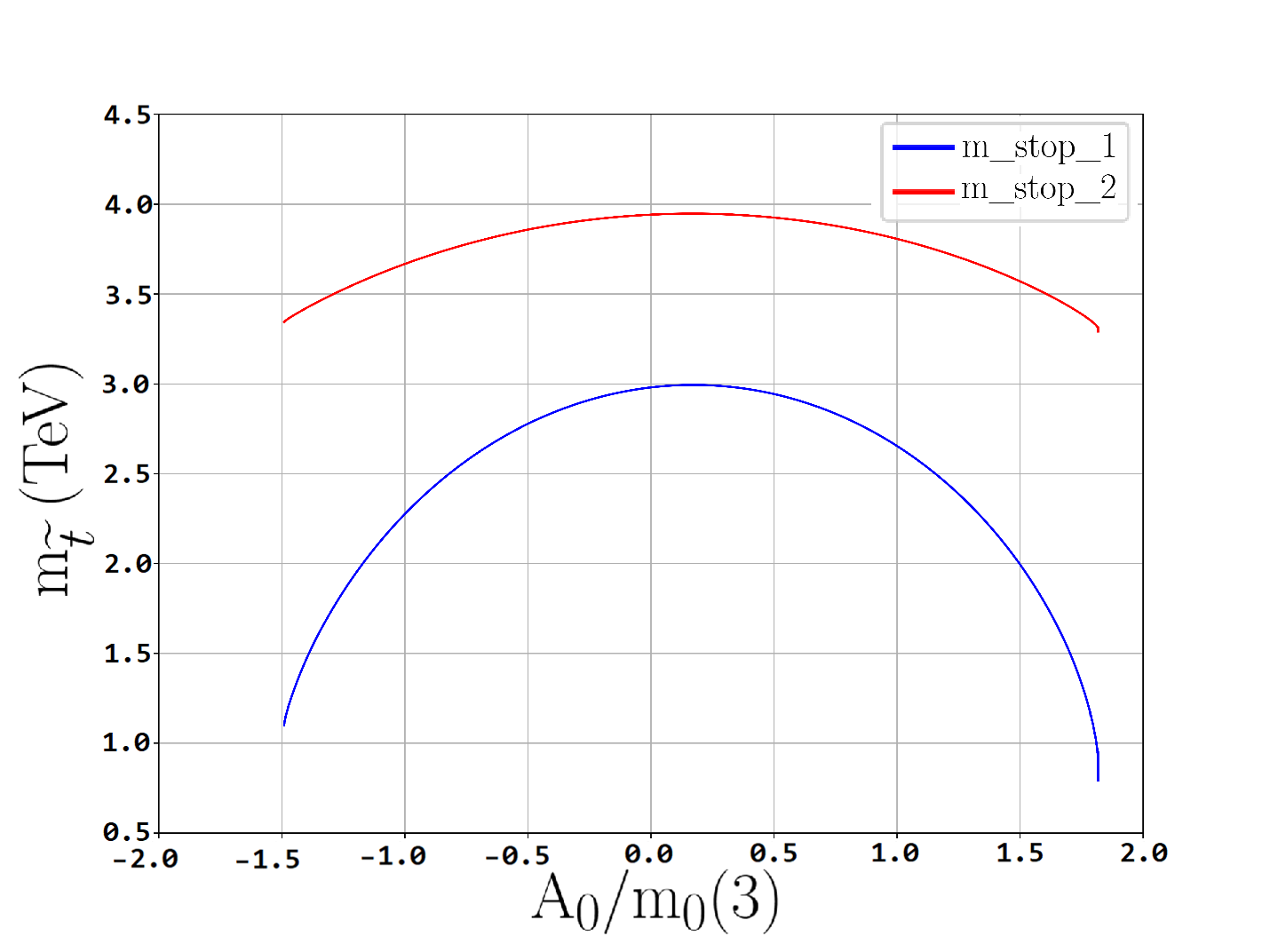}\\
\includegraphics[height=0.28\textheight]{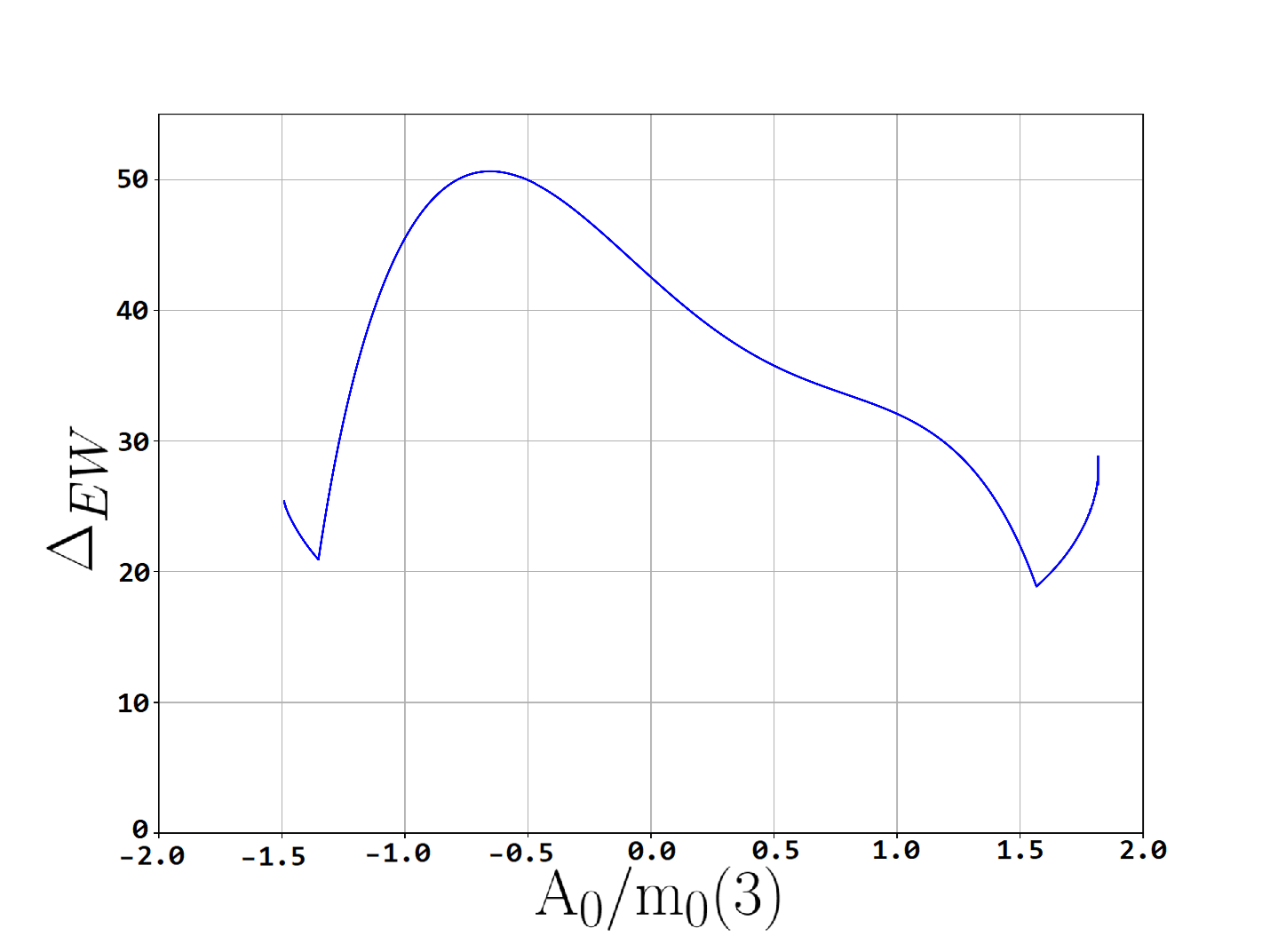}
\includegraphics[height=0.28\textheight]{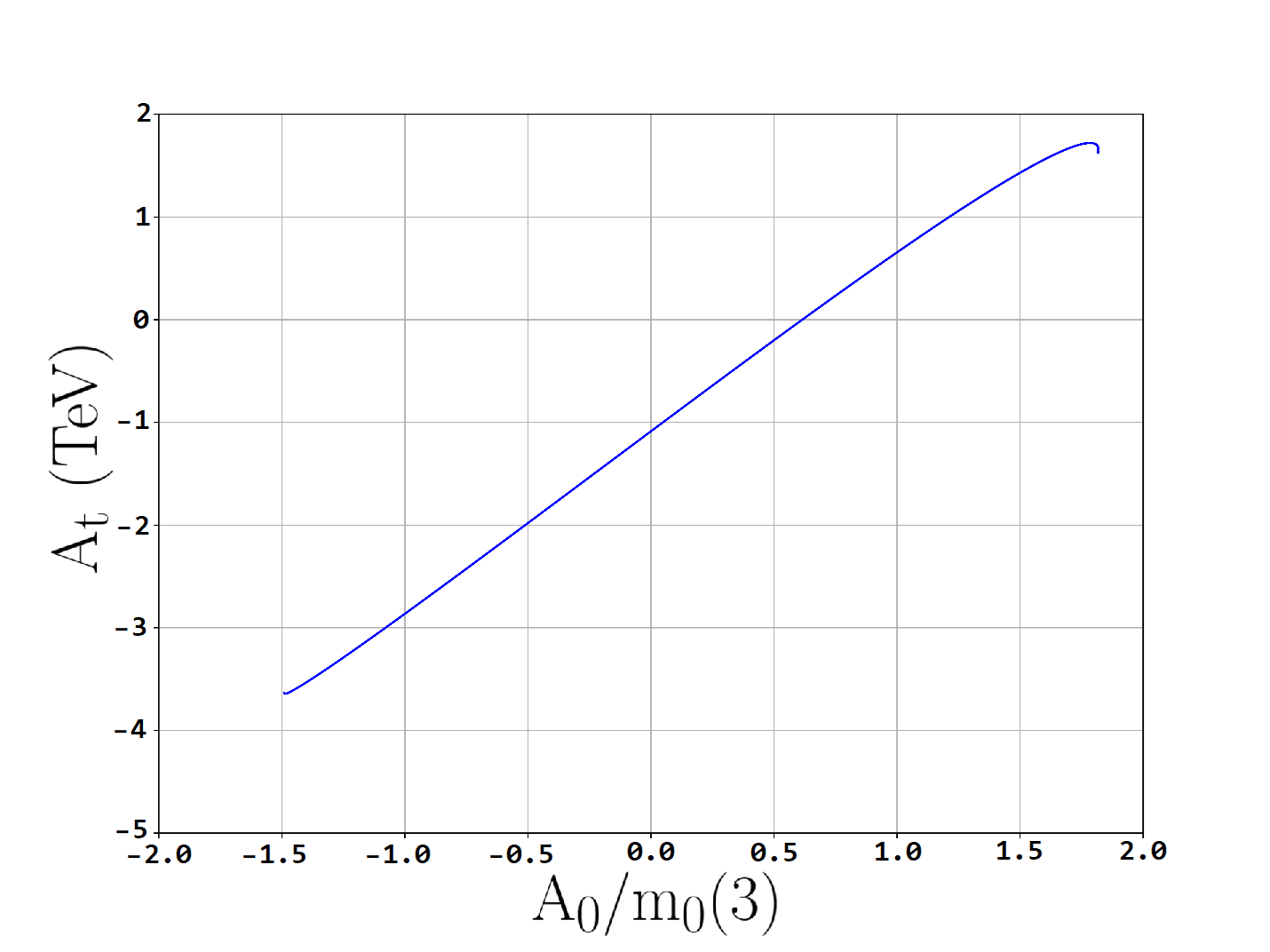}
\caption{Values of {\it a}) $m_h$, {\it b}) $m_{\tst_{1,2}}$, 
{\it c}) $\Delta_{EW}$ and {\it d}) $A_t(Q)$ vs. $A_0/m_0(3)$
for the NUHM3 model with $m_0(1,2)=10$ TeV, $m_0(3)=5$ TeV, 
$m_{1/2}=0.7$ TeV, $\tan\beta =10$ with $\mu=200$ GeV and
$m_A=2$ TeV.
\label{fig:A0m0}}
\end{center}
\end{figure}

In Fig. \ref{fig:A0m0}, we show the values of {\it a}) $m_h$, {\it b}) $m_{\tst_{1,2}}$ 
,
{\it c}) $\Delta_{EW}$ and {\it d}) $A_t$ versus $A_0/m_0$ for the NUHM3
model with parameters as in the caption but with varying $A_0$. 
(NUHM3 splits first/second generation sfermion soft terms from third generation ones
so that $m_0(1,2)\ne m_0(3)$.)
These plots are obtained using SOFTSUSY and
can be compared to similar plots in Ref. \cite{Baer:2012up} using Isasugra.
We see from frame {\it a}) that the value of $m_h$ is actually maximal
at large negative $A_t$ values (which are shown in frame {\it d})).
The large mixing in the stop sector lifts the value of $m_h$ to the 125 GeV 
regime, but in this case only for {\it negative} $A_t$ values. The stop mass eigenstates
are shown in frame {\it b}) where again, when there is large mixing, the eigenstates
have the largest splittings and $m_{\tst_1}$ becomes lowest in value.
In frame {\it c}), we show the corresponding value of $\Delta_{EW}$. 
Here we see that for large trilinear $A_t$, then there can be large
cancellations in $\Sigma_u^u(\tst_{1,2})$ which lead to decreased finetuning.
The kinks in the curve occur due to transitions from one maximal
contribution to $\Delta_{EW}$ to a different one. The dominant contributions to
$\Delta_{EW}$ in the middle of the plot comes from top-squark
contributions whilst the left and right edges come from tau-slepton
contributions (as in Fig. 2 of Ref. \cite{Baer:2012up}).
The low value of $\Delta_{EW}$ coincides with the uplift in $m_h$ to $\sim 125$ GeV
for large negative values of $A_t$.

\begin{figure}[!htbp]
\begin{center}
\includegraphics[height=0.5\textheight]{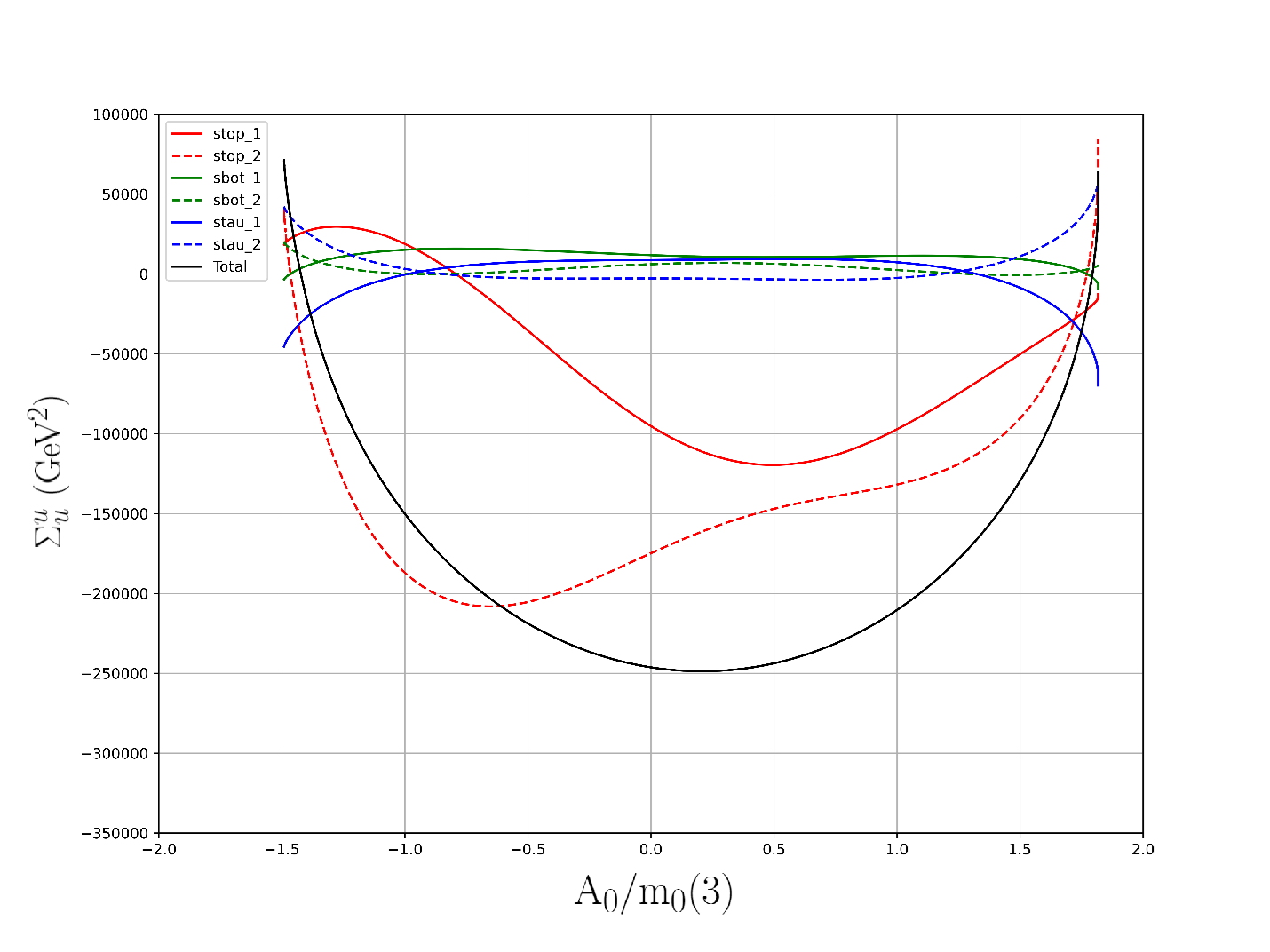}
\caption{Third generation contributions to $\Delta_{EW}$ for the same
model parameters as Fig. \ref{fig:A0m0} vs. $A_0/m_0(3)$
in the $m_0$ vs. $m_{1/2}$ plane of the NUHM2 model with 
$\mu =200$, $\tan\beta =10$, $A_0=-1.6 m_0$ and $m_A=2$ TeV.
\label{fig:Sig3}}
\end{center}
\end{figure}

In Fig. \ref{fig:Sig3}, we show the third generation contributions to 
$\Delta_{EW}$ vs. $A_0/m_0(3)$ for the same parameters as in Fig. \ref{fig:A0m0}, 
but using SOFTSUSY. These can be compared with the same plot using Isasugra 
in Fig. 2 of Ref. \cite{Baer:2012up}. Here, we see that the contributions from
staus and sbottoms are generally rather small, and the top-squark contributions
typically dominate. But for large $|A_0/m_0(3)|$, then cancellations in
both $\Sigma_u^u(\tst_1)$ and $\Sigma_u^u(\tst_2)$ occur, and the stop contributions 
become comparable to those of the other third generation sparticles, giving
reduced finetuning and greater naturalness.
\subsection{Natural regions of $m_0$ vs. $m_{1/2}$ plane}
\label{ssec:planes}

\begin{figure}[!htbp]
\begin{center}
\includegraphics[height=0.5\textheight]{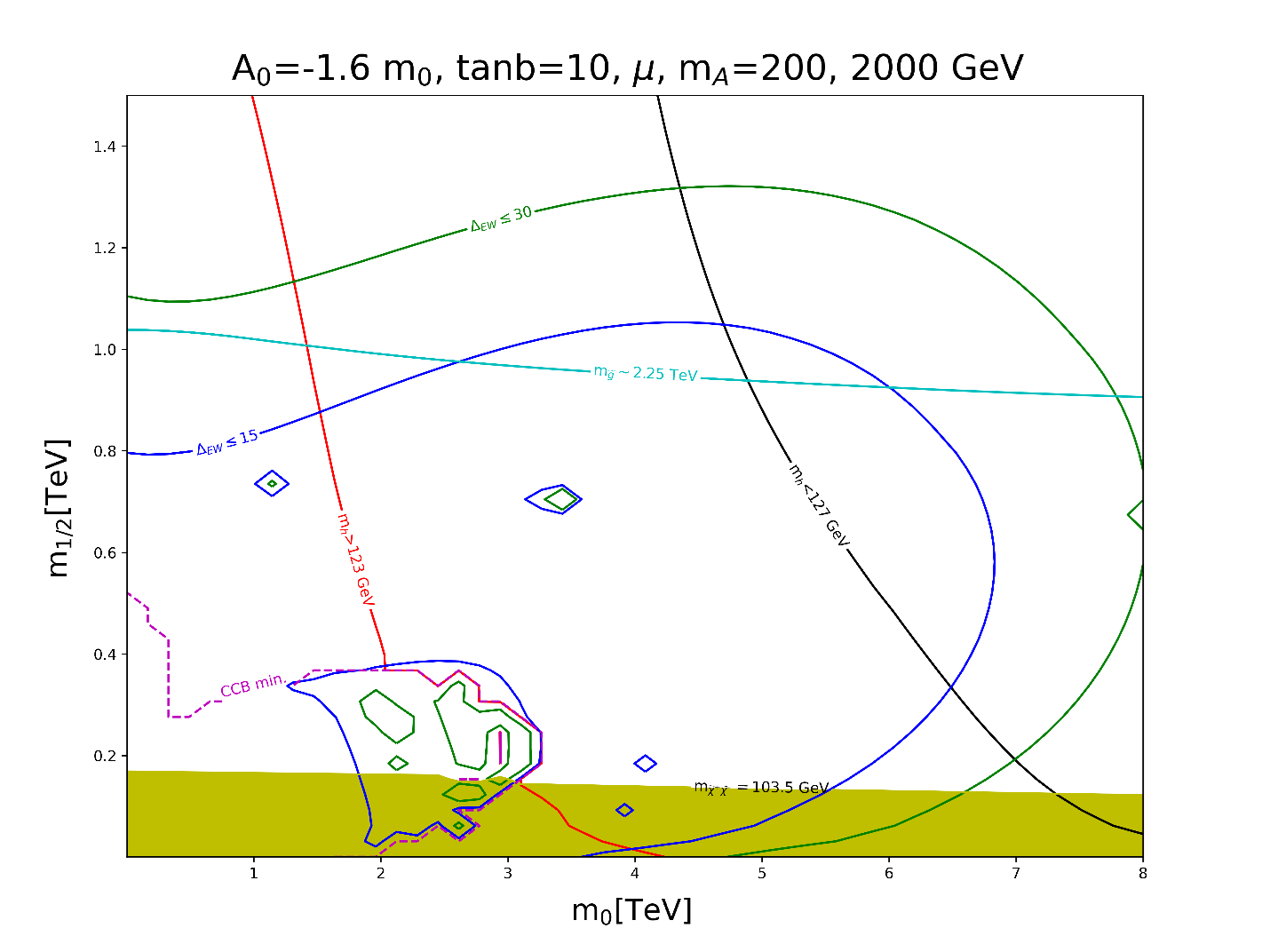}
\caption{Contours of naturalness measure $\Delta_{EW}$ and $m_h$
in the $m_0$ vs. $m_{1/2}$ plane of the NUHM2 model with 
$\mu =200$, $\tan\beta =10$, $A_0=-1.6 m_0$ and $m_A=2$ TeV.
\label{fig:nuhm2}}
\end{center}
\end{figure}

In Fig. \ref{fig:nuhm2}, we show the $m_0$ vs. $m_{1/2}$ parameter plane for
the NUHM2 model with $A_0=-1.6 m_0$, $\mu =200$ GeV and $m_A=2$ TeV. 
The plot is generated using SOFTSUSY but can be compared with similar results
from Isasugra in Fig. 8{\it b} of Ref. \cite{Baer:2019cae}. From the plot, we see the
lower-left corner is actually excluded due to charge-or-color-breaking (CCB) vacua
which occur for too large $A_0$ values. Both SOFTSUSY and Isasugra generate CCB regions
there. We also show contours of Higgs mass $m_h=123$ and $127$ GeV. These are
qualitatively similar to the Isasugra results but shifted to the right by a couple hundred
GeV in $m_0$. Thus, much of the parameter space allows for the measured Higgs mass
$m_h\sim 125$ GeV. We also show naturalness contours for $\Delta_{EW}=15$ and 30.
These can also be compared against the LHC Run 2 gluino mass limit 
$m_{\tg}\agt 2.25$ TeV as shown by the light blue contour.
The important point is that both SOFTSUSY and Isasugra agree that the bulk of this parameter space plane is 
EW natural, in accord with LHC gluino mass limits, and in accord with the 
measured Higgs mass. This is in contrast to older naturalness measures 
which required much lower gluino masses\cite{Ellis:1986yg,Barbieri:1987fn,Dimopoulos:1995mi,Anderson:1994tr} 
and also Higgs boson masses\cite{Cassel:2009cx}.

\section{String landscape distributions from SOFTSUSY}
\label{sec:landscape}

In this section, we wish to compare SUSY landscape predictions using a spectrum
calculator other than Isasugra. Here, we choose SOFTSUSY.
The assumption is that the MSSM is the low energy EFT in a fertile patch of landscape
vacua, but with different sets of soft SUSY breaking terms in each pocket universe, and
hence a different value for the weak scale $m_{weak}^{PU}\ne m_{weak}^{OU}$ in
each pocket universe (here, $OU=$ our universe). Following Douglas\cite{Douglas:2004qg}, Susskind\cite{Susskind:2004uv} and
Arkani-Hamed, Dimopoulos and Kachru\cite{Arkani-Hamed:2005zuc}, we will assume the soft terms scan in 
the landscape as a power-law: $m_{soft}^n$ where $n=2n_F+n_D-1$ with 
$n_F$ the number of $F$ breaking fields and $n_D$ the number of $D$ breaking fields. 
Here, we assume $n=1$ corresponding to SUSY breaking by a single $F$ term, where
$F$ is distributed as a random complex number. As in Ref. \cite{Baer:2020vad}, we 
expect each soft term in the NUHM3 model to scan independently.

We perform the linear soft term scan over NUHM3 space as follows:
\bi
\item $m_0(1,2):\ 0.1-60$ TeV,
\item $m_0(3):\ 0.1-20$ TeV,
\item $m_{1/2}:\ 0.5-10$ TeV,
\item $A_0:\ -50\ -\ 0$ TeV,
\item $m_A:\ 0.3-10$ TeV
\ei
with $\mu=200$ GeV and $\tan\beta$ scanned uniformly between $3-60$.
The goal is to set upper limits on scan parameters that are beyond the 
upper limits that will result from imposing the anthropic conditions.
We also require appropriate EWSB and so veto vacua with CCB minima or with 
no EWSB. We also require the Agrawal condition on the magnitude of the weak scale\cite{Agrawal:1997gf}: 
\bi
\item $m_Z^{PU}<4 m_Z^{OU}$
\ei
which corresponds to $\Delta_{EW}<30$. Thus, the following results 
generated using SOFTSUSY can be compared to comparable results in Ref. \cite{Baer:2017uvn}
using Isasugra.

\begin{figure}[!htbp]
\begin{center}
\includegraphics[height=0.25\textheight]{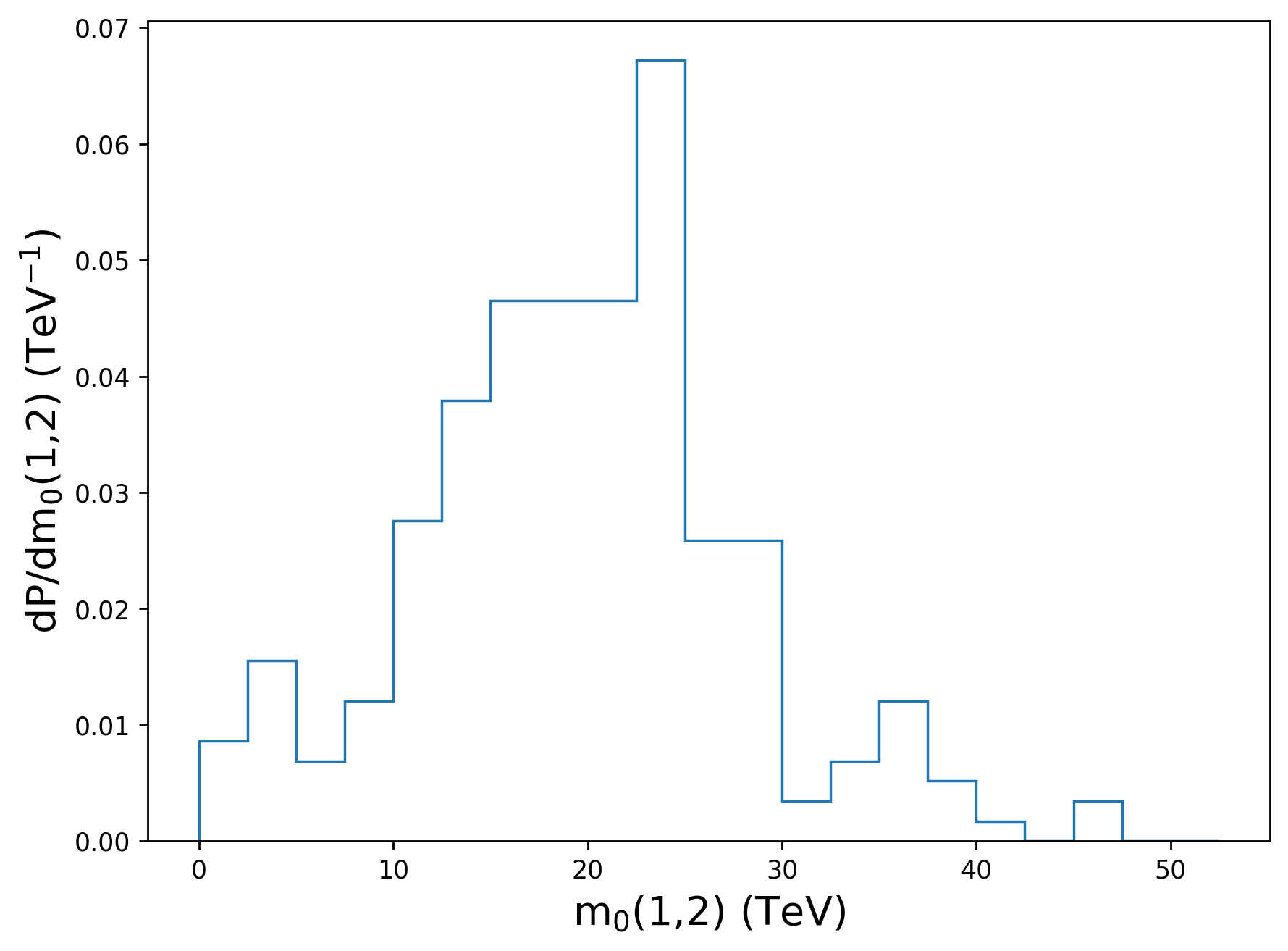}
\includegraphics[height=0.25\textheight]{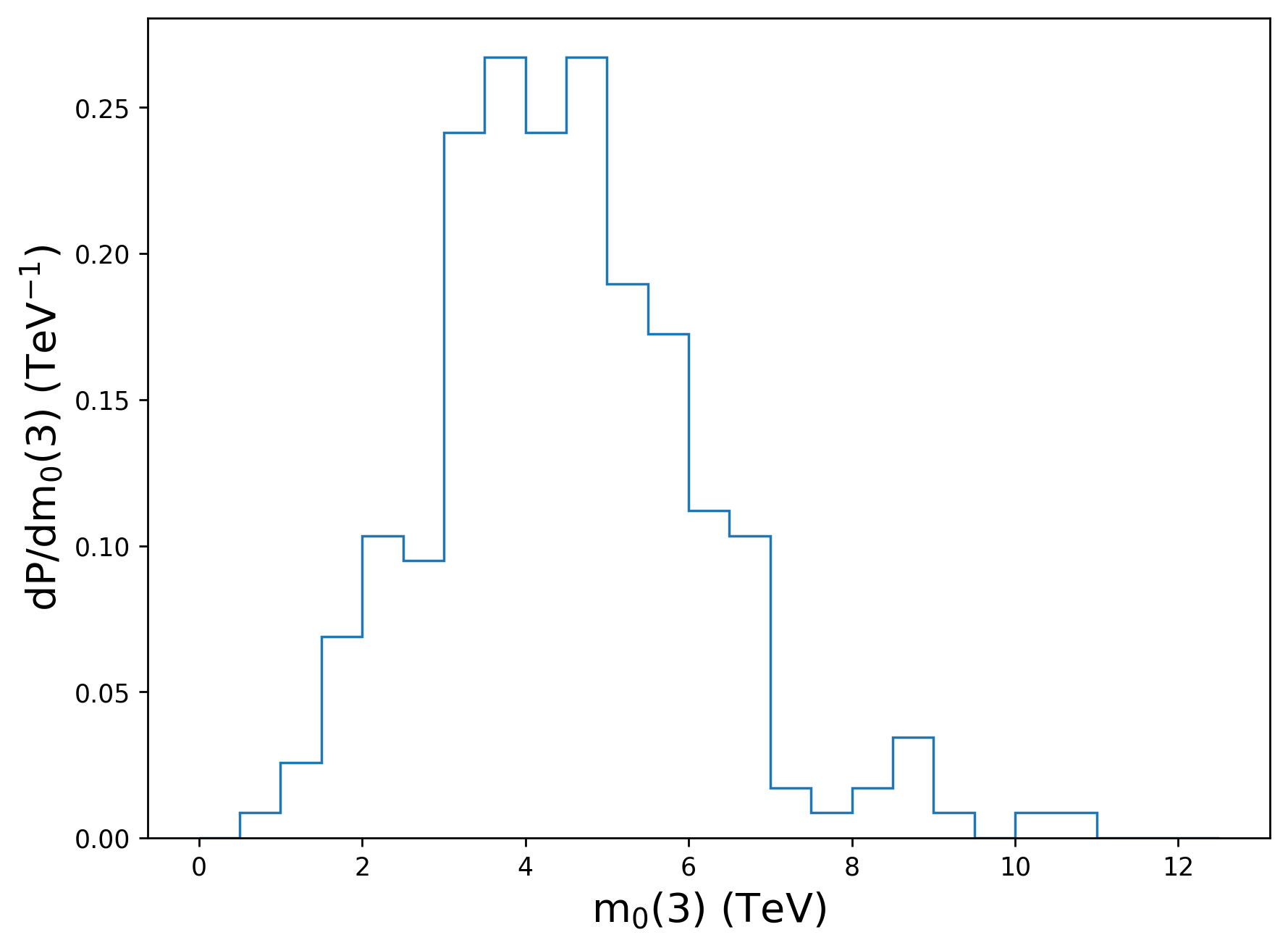}\\
\includegraphics[height=0.25\textheight]{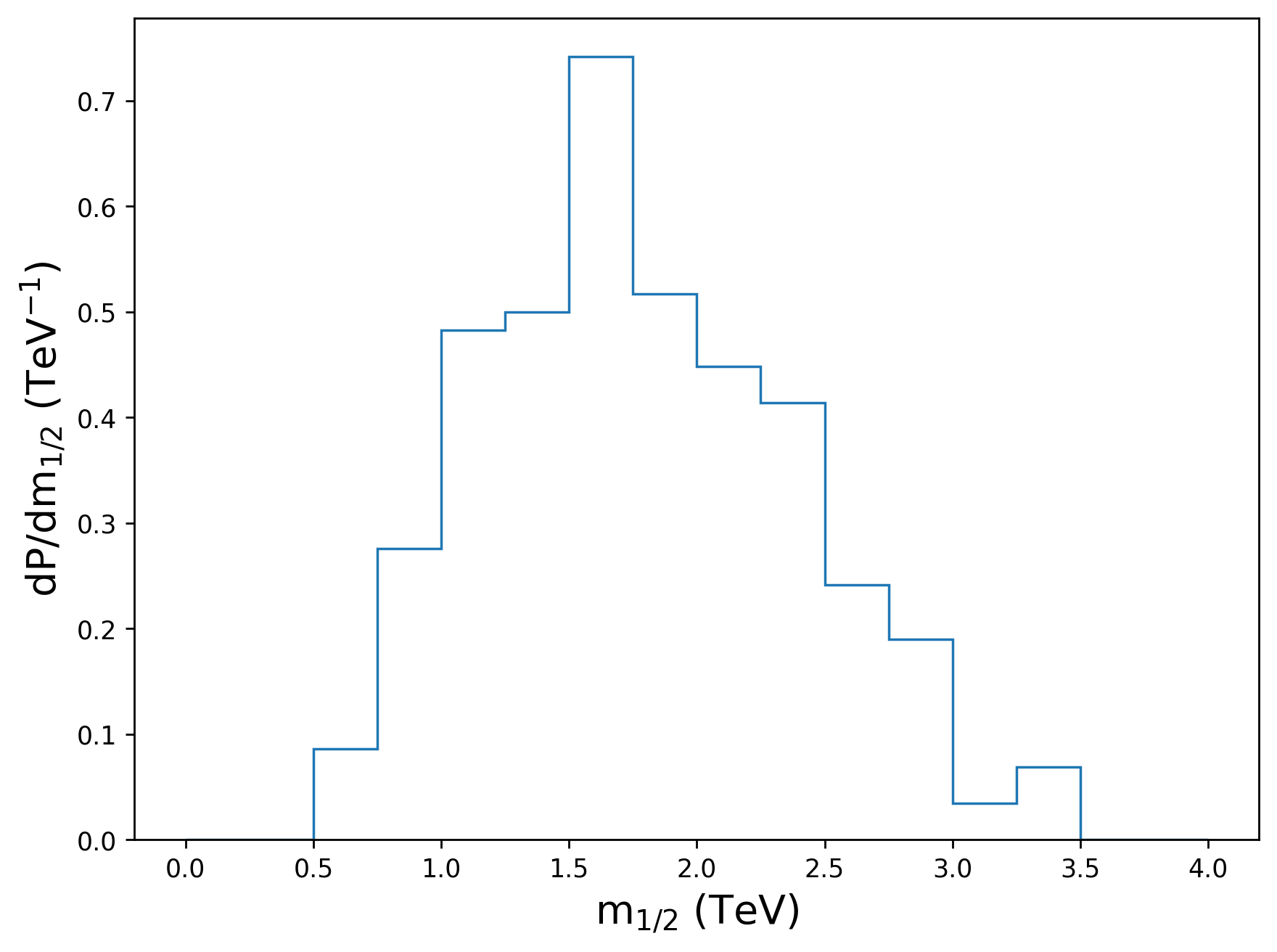}
\includegraphics[height=0.25\textheight]{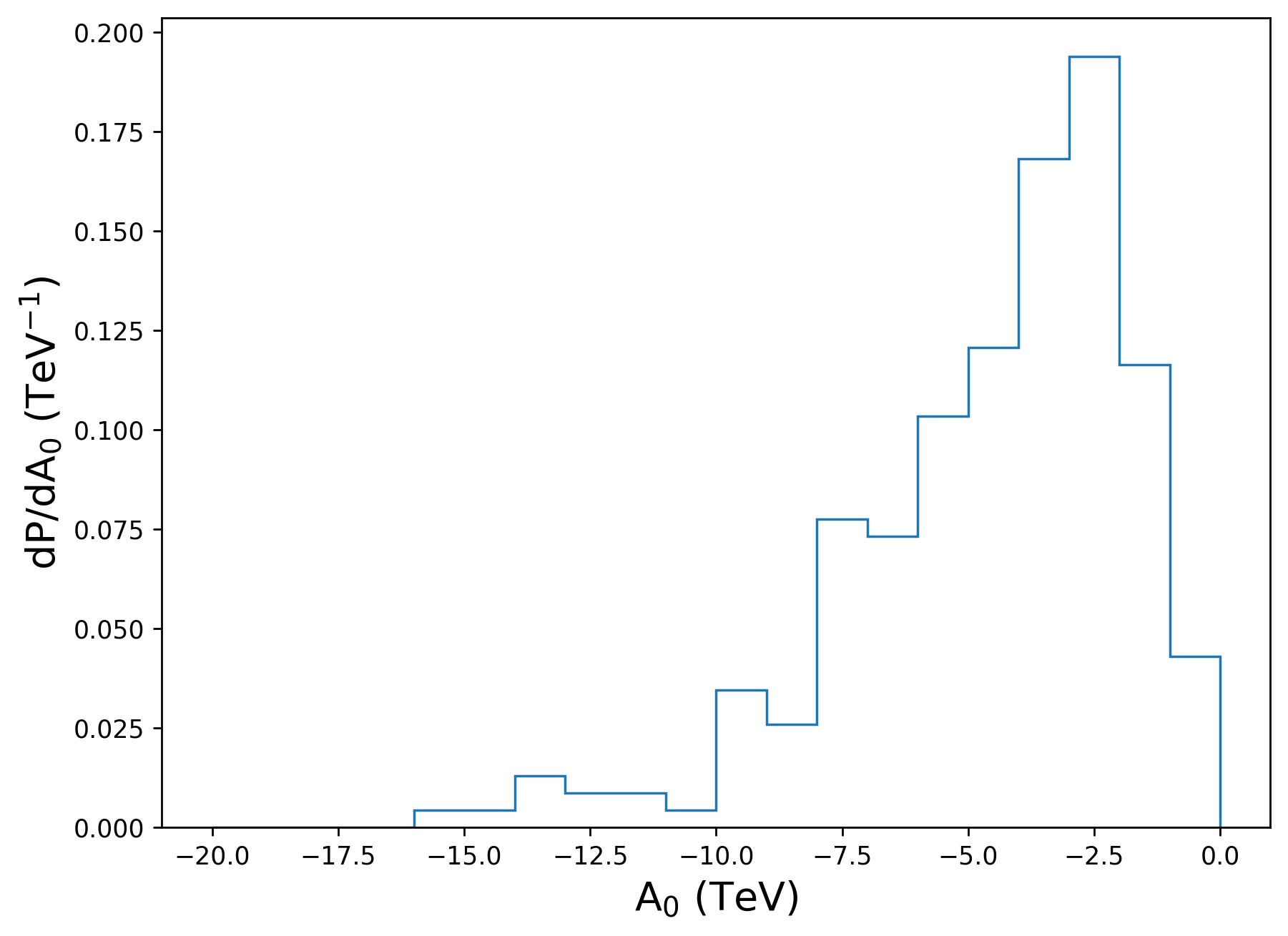}
\caption{Distribution of soft SUSY breaking terms in subset of
landscape vacua with the NUHM3 model as low energy EFT and an 
$n=1$ statistical draw.
We show distributions in {\it a}) $m_0(1,2)$, {\it b}) $m_0(3)$, 
{\it c}) $m_{1/2}$ and {\it d}) $A_0$.
\label{fig:land1}}
\end{center}
\end{figure}

The resultant distribution in input parameters from the SOFTSUSY scan
is shown in Fig. \ref{fig:land1}. In frame {\it a}), we see the distribution in first/second generation GUT scale soft masses peaks around 20 TeV and spans $\sim 5-40$ TeV.
While first/second generation scalars contribute to the weak scale via
Yukawa suppressed terms, they also contribute via EW $D$-term contributions
(which largely cancel due to cancellation of EW quantum numbers) and via
two-loop RG terms which, when large, drive third generation scalars to tachyonic 
values\cite{Baer:2000xa}. The last of these effectively sets the upper bound, allowing for
$m_0(1,2)$ as high as $40-50$ TeV. This provides a mixed decoupling/quasi-degeneracy
solution to the SUSY flavor and CP problems\cite{Baer:2019zfl} since the upper bound is 
flavor independent. In frame {\it b}), the third generation soft masses are bounded by much lower values: $m_0(3)\sim 1-10$ TeV with a peak around 5 TeV. Here, the upper bound
comes from requiring not-too-large values of $\Sigma_u^u(\tst_{1,2})$ values. In frame {\it c}), we see the distribution in $m_{1/2}$, which ranges form $0.5-3$ TeV. The upper bound is set because if $m_{1/2}$ is too large, it drives the stop soft terms to large
values and again $\Sigma_u^u(\tst_{1,2})$ gets too big. In frame {\it d}), we plot the distribution in $-A_0$. There is hardly any probability around $A_0\sim 0$ so we 
expect large mixing in the stop sector, which ends up driving $m_h$ to large values.
But $A_0$ cannot become too large (negative) lest it pushes the top squark soft terms to tachyonic values via RG running.

\begin{figure}[!htbp]
\begin{center}
\includegraphics[height=0.25\textheight]{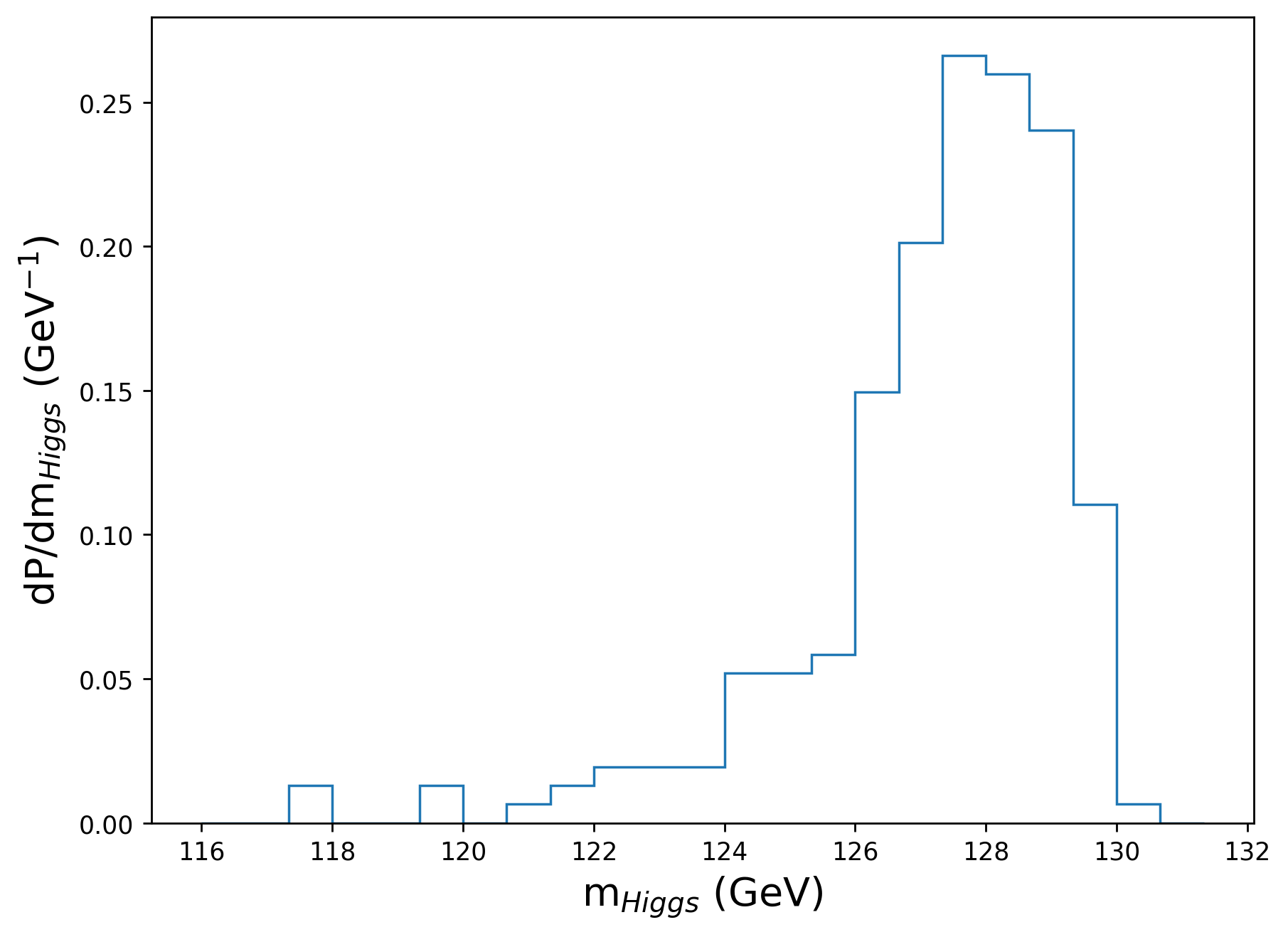}
\includegraphics[height=0.25\textheight]{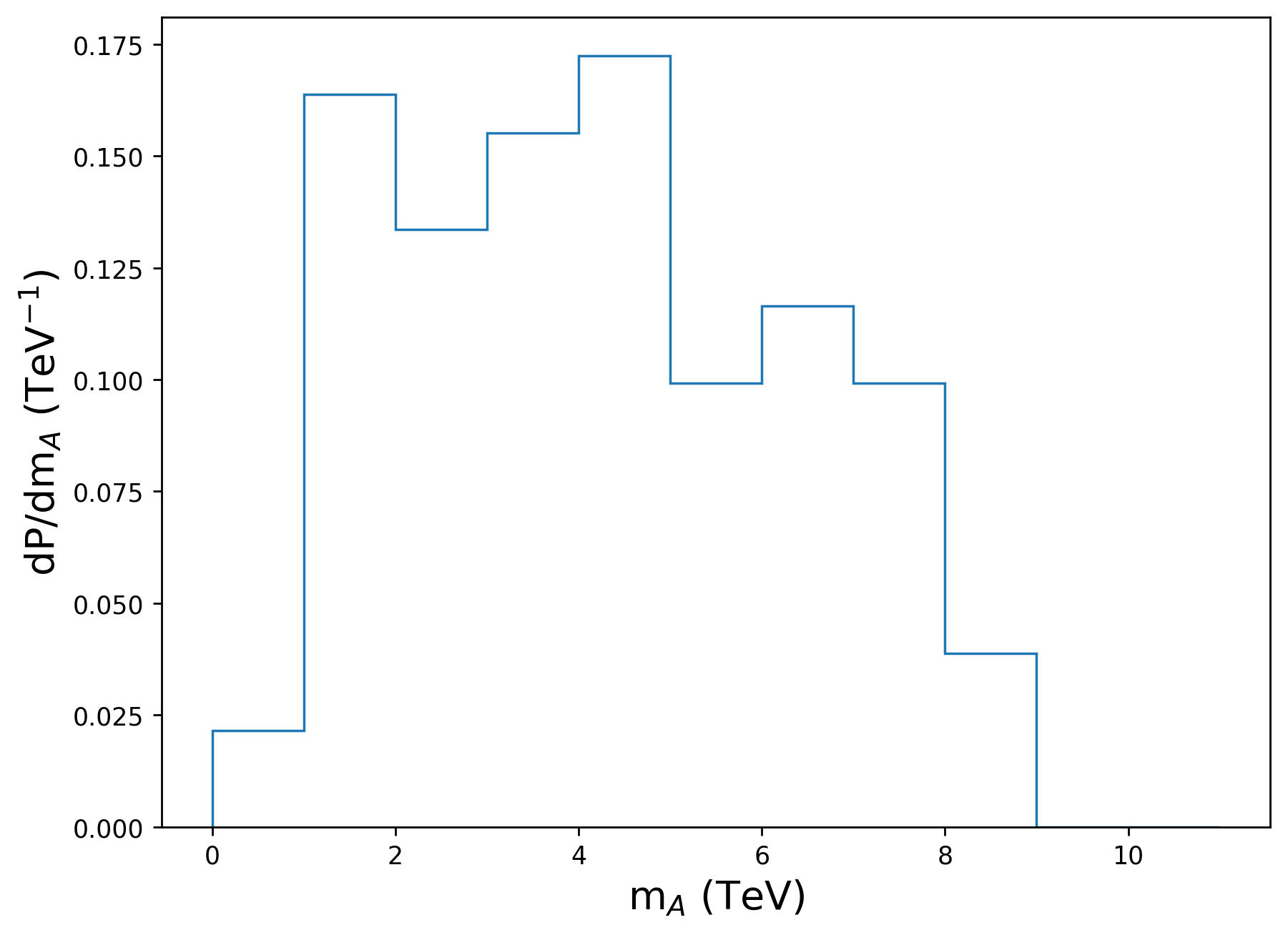}\\
\includegraphics[height=0.25\textheight]{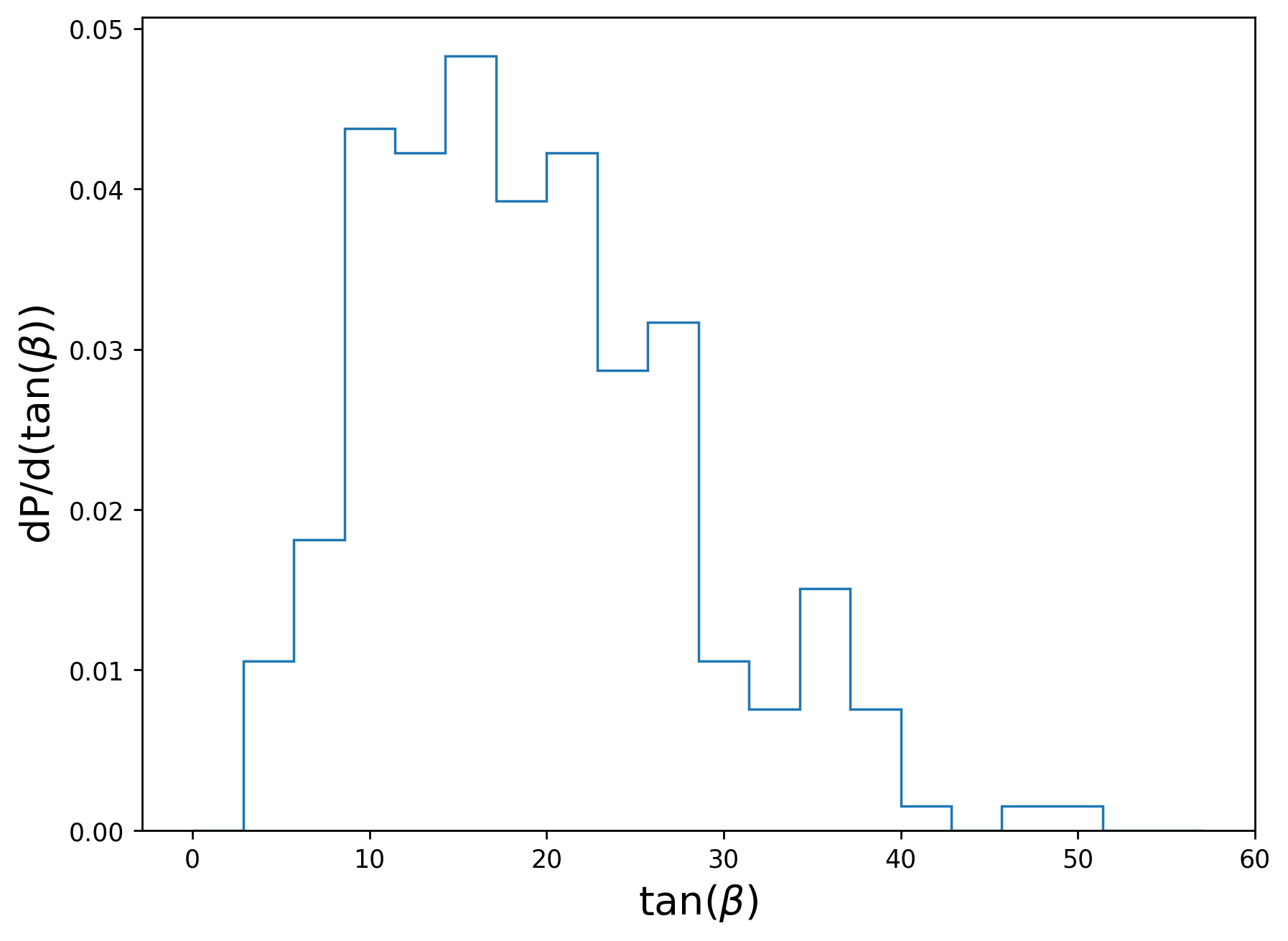}
\includegraphics[height=0.25\textheight]{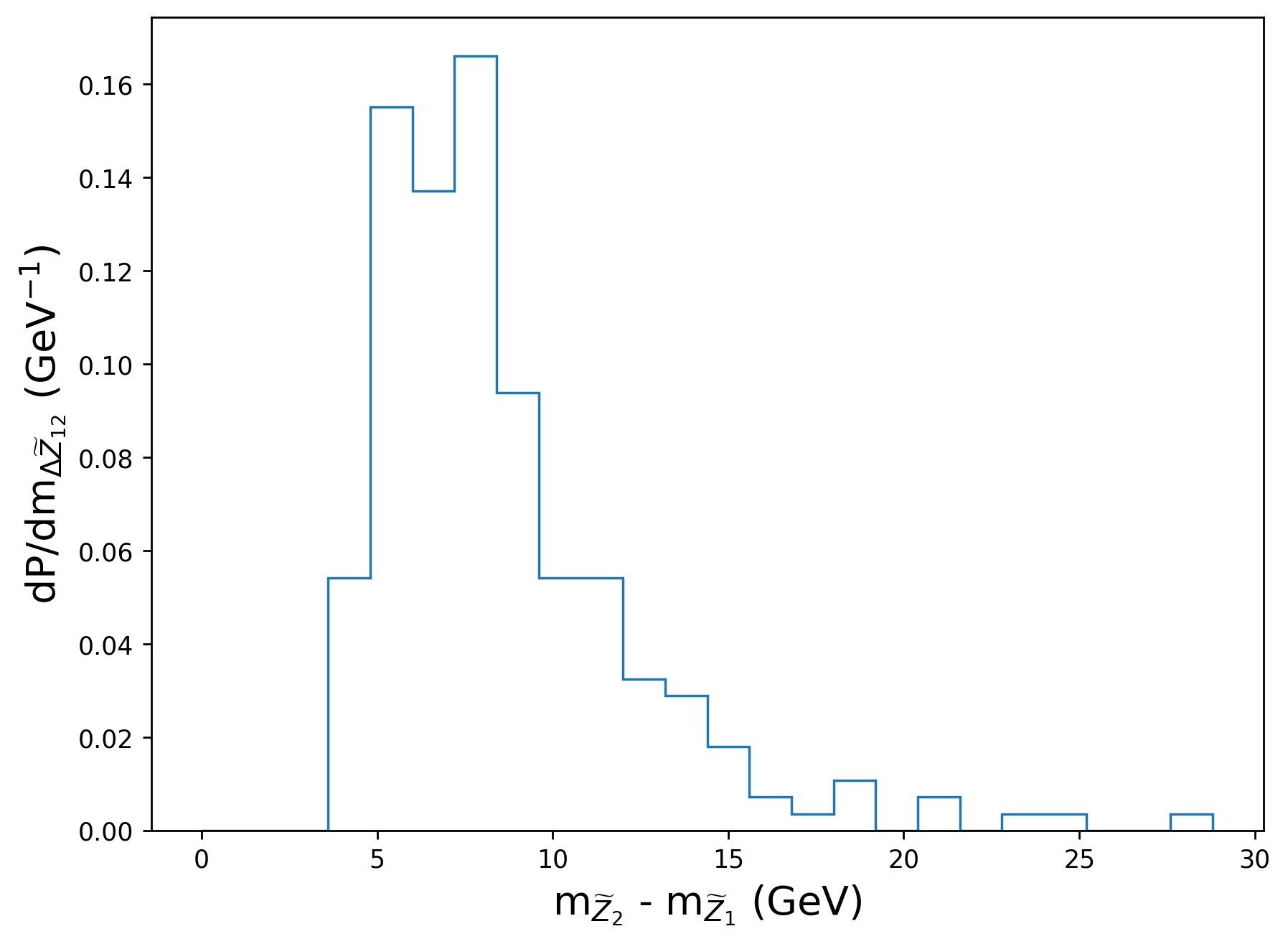}
\caption{Distribution of Higgs and EWino sector from a  subset of
landscape vacua with NUHM3 model as low energy EFT and an 
$n=1$ statistical draw.
We show distributions in {\it a}) $m_h$, {\it b}) $m_A$, 
{\it c}) $\tan\beta$ and {\it d}) $m_{\tchi_2^0}-m_{\tchi_1^0}$.
\label{fig:land2}}
\end{center}
\end{figure}

In Fig. \ref{fig:land2}, we show $n=1$ landscape scan probability 
distributions from the EW sector. In frame {\it a}), we show the distribution
in light Higgs mass $m_h$. Using SOFTSUSY, the distribution rises to a peak
$m_h\sim 128$ GeV, which is several GeV higher than the result from Isasugra.
This is consistent with SOFTSUSY generating $m_h$ typically a couple GeV higher
than Isasugra. In frame {\it b}), we see the distribution in $m_A$ which runs
from $1-9$ TeV with a peak around 4 TeV. Thus, we expect a decoupled SUSY 
Higgs sector with the couplings of $h$ being very close to their SM values.
In frame {\it c}), the distribution in $\tan\beta$ peaks around $10-20$.
The upper bound is set because if $\tan\beta$ gets too big, then the
$\Sigma_u^u(\tb_{1,2})$ terms become large (large $b$ and $\tau$ Yukawa couplings)
and the model is more likely to generate a large $m_{weak}^{PU}$.
In frame {\it d}), we show the $\tchi_2^0-\tchi_1^0$ mass difference which is important
for LHC higgsino-pair searches\cite{Baer:2014kya}. In this case, the landscape predicts
$m_{\tchi_2^0}-m_{\tchi_1^0}\sim 5-15$ GeV.

\begin{figure}[!htbp]
\begin{center}
\includegraphics[height=0.25\textheight]{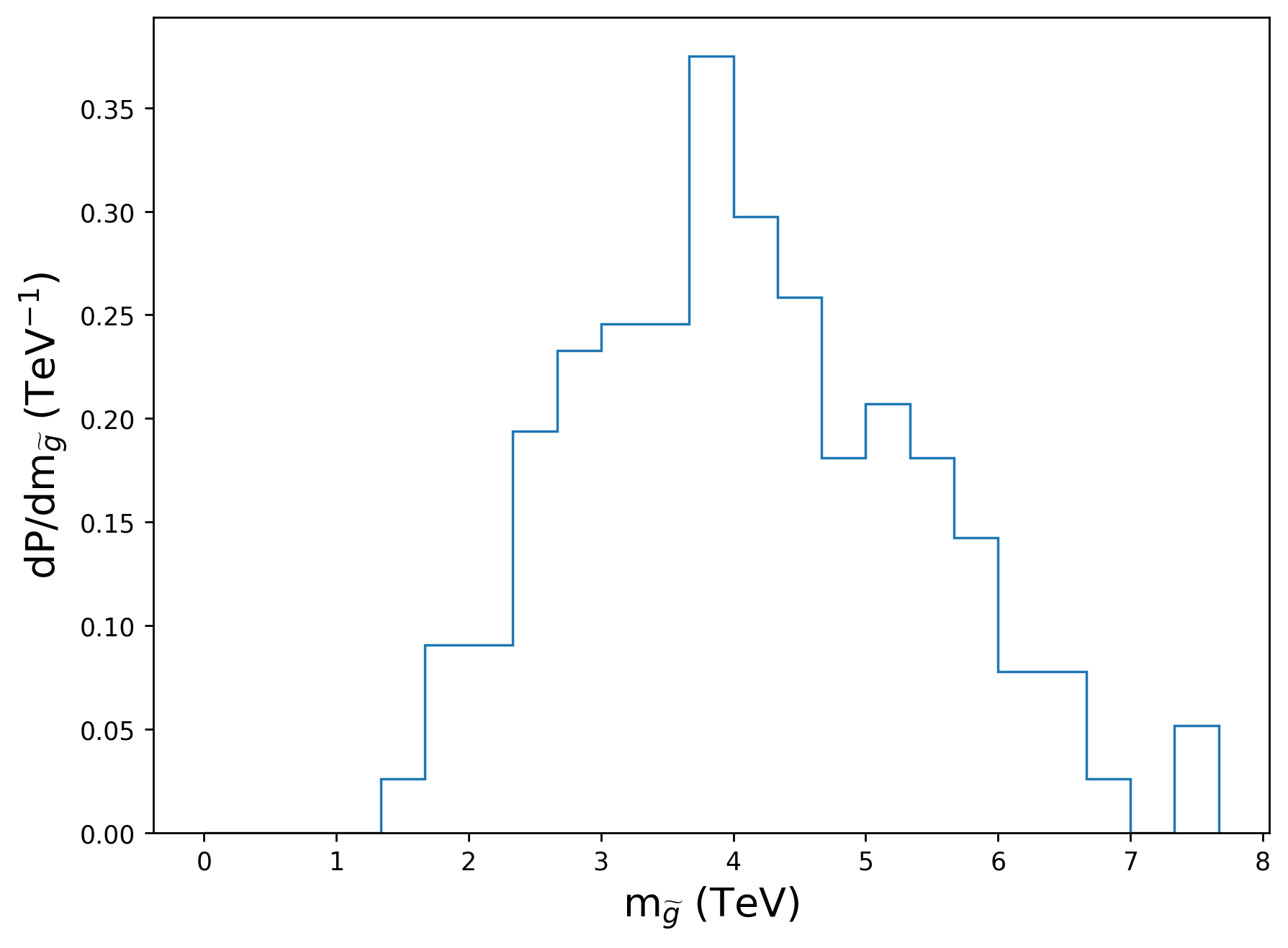}
\includegraphics[height=0.25\textheight]{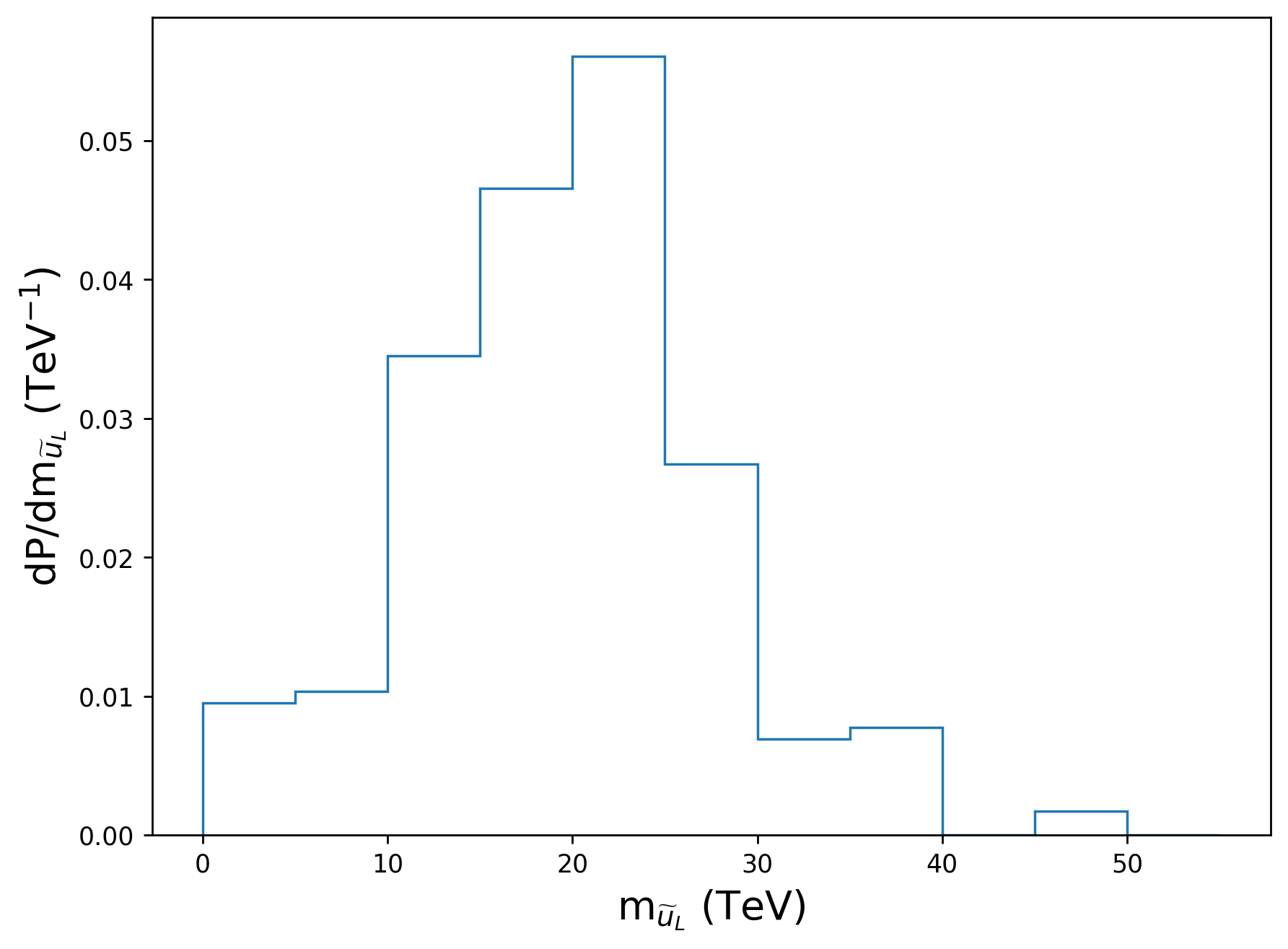}\\
\includegraphics[height=0.25\textheight]{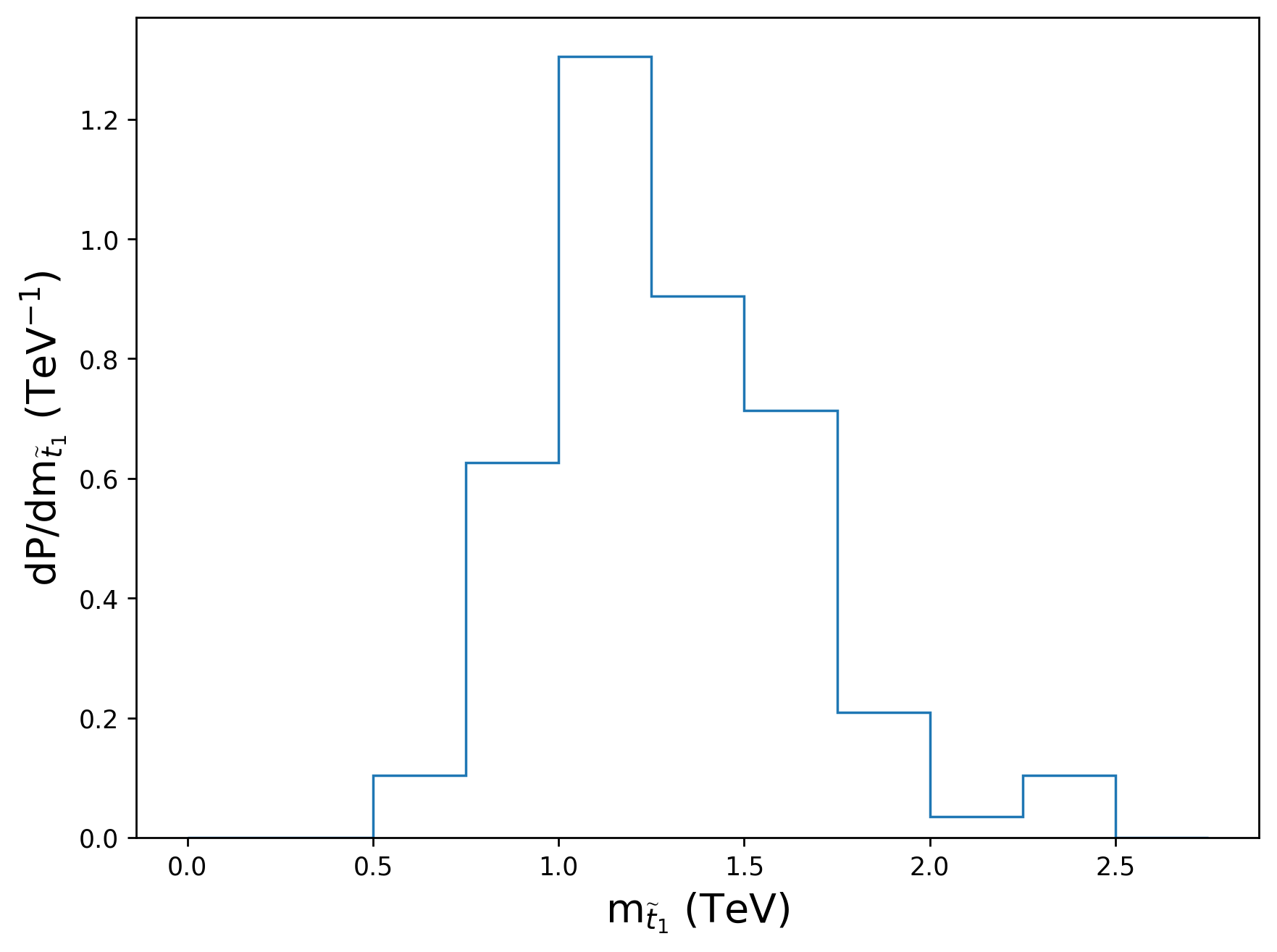}
\includegraphics[height=0.25\textheight]{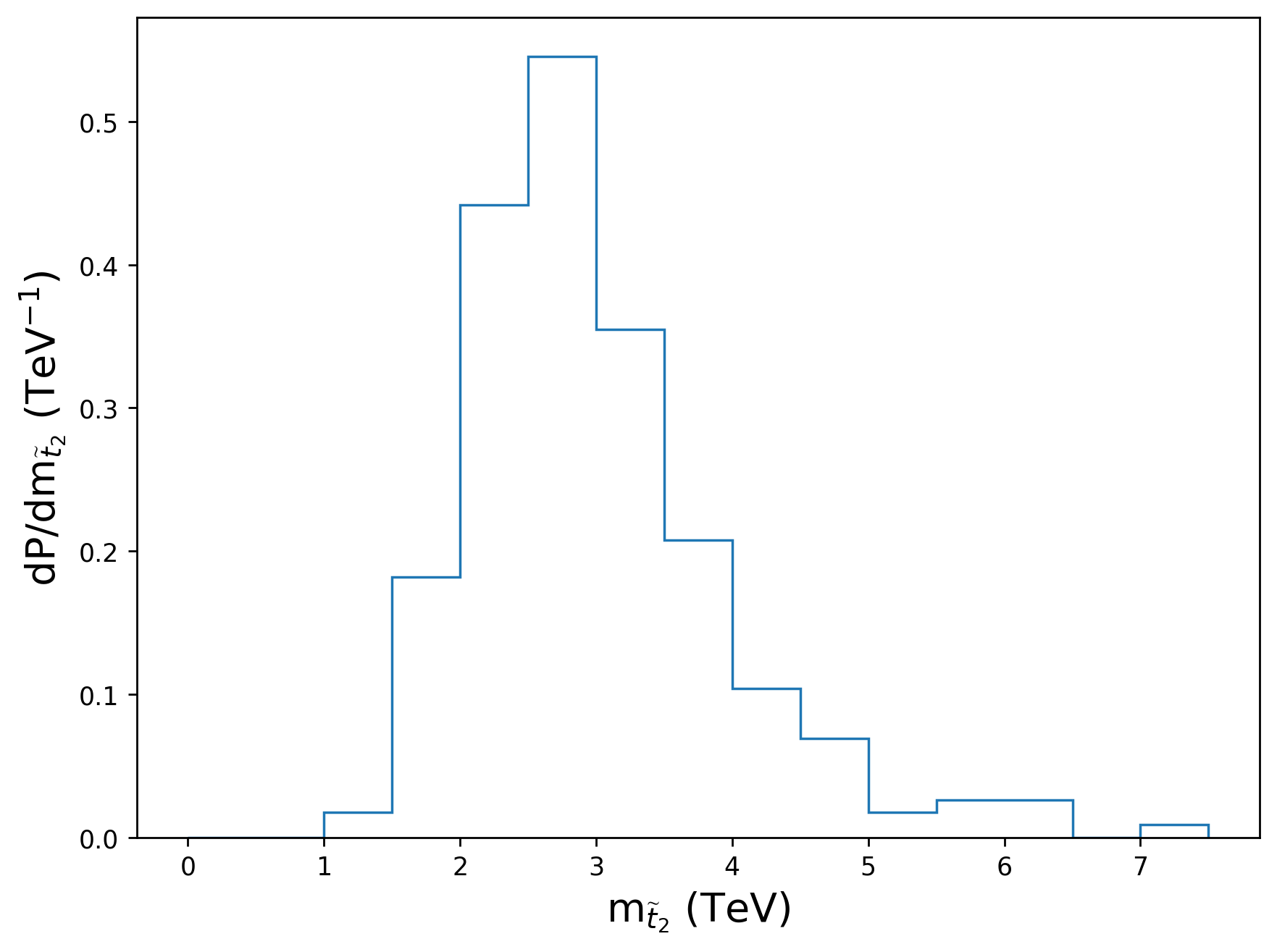}
\caption{Distribution of strongly interacting sector masses from a  subset of
landscape vacua with NUHM3 model as low energy EFT and an 
$n=1$ statistical draw.
We show distributions in {\it a}) $m_{\tg}$, {\it b}) $m_{\tu_L}$, 
{\it c}) $m_{\tst_1}$ and {\it d}) $m_{\tst_2}$.
\label{fig:land3}}
\end{center}
\end{figure}

In Fig. \ref{fig:land3}, we show $n=1$ landscape distributions for strongly interacting
SUSY particles using SOFTSUSY. In frame {\it a}), we see the gluino mass
$m_{\tg}\sim 2-7$ TeV. The LHC13 limit of $m_{\tg}\agt 2.25$ TeV just excludes the
lower edge of the expected values. Thus, from the landscape point of view, it is no surprise that LHC has so far failed to detect gluinos. In frame {\it b}), we 
show the distribution in left up-squark mass (which is indicative of both first and second
generation sfermion masses). The distribution ranges from $m_{\tu_L}\sim10-40$ TeV, 
so these sparticles are likely far beyond LHC reach. In frame {\it c}), we show
the distribution in light top-squark mass. Here we find $m_{\tst_1}\sim 1-2$ TeV, so again
it may come as no surprise that LHC has so far not discovered evidence of top-squark 
pair production. The heavier top squark distribution $m_{\tst_2}$ is shown in 
frame {\it d}). Here we see it ranges between $2-5$ TeV.
All these results are in qualitative agreement with previous results generated
using Isasugra\cite{Baer:2017uvn}.

\section{Conclusions}
\label{sec:conclude}

We have created a publicly available computer code DEW4SLHA which computes the
electroweak finetuning measure $\Delta_{EW}$ from any SUSY/Higgs spectrum 
generator which produces the standard SUSY Les Houches Accord output file.
The code then allows us to compare naturalness and landscape predictions from various spectra codes against each other.

We have used the DEW4SLHA code in Sec. \ref{sec:BMpoints} to compare natural SUSY
spectra from Isasugra, SOFTSUSY, SUSPECT and SPHENO. The outputs for a natural SUSY
benchmark point are generally in good agreement although SOFTSUSY and SUSPECT gives
values of $m_h$ a couple GeV higher than Isasugra or SPHENO. We also computed the top
44 contributions to $\Delta_{EW}$ which are typically in agreement although SPHENO
generates slightly more stop mixing than the other codes. The SOFTSUSY code was
used to display cancellations in $\Sigma_u^u(\tst_{1,2})$ that lead to increased
naturalness for large stop mixing (which also lifts the value of 
$m_h\sim 125$ GeV). 

We also used SOFTSUSY to corroborate predictions for sparticle and Higgs boson masses 
from the string landscape where a draw to large soft terms along with an anthropic
requirement on the weak scale $m_{weak}^{PU}<4 m_{weak}^{OU}$ leads to statistical
predictions from compactified string models with the MSSM as the low energy EFT.
SOFTSUSY generates a Higgs mass peak $m_h\sim 127-128$ GeV, slightly higher than 
Isasugra. SOFTSUSY also generates sparticle mass spectra typically beyond LHC13 reach,
confirming earlier Isasugra results.

\section*{Acknowledgements} 

This work has been performed as part of a contribution to the Snowmass 2022
workshop.
This material is based upon work supported by the U.S. Department of Energy, 
Office of Science, Office of Basic Energy Sciences Energy Frontier Research 
Centers program under Award Number DE-SC-0009956 and U.S. Department of Energy 
Grant DE-SC-0017647. 

\appendix
\label{appendix}

\numberwithin{equation}{section}
\section{Appendix of corrections}
\label{sec:appendix}
	Starting with the effective potential, we have to one-loop order
	\myeq
	V_{\text{Higgs}}&=V_{\text{tree}}+\Delta V
	\myeqend
	
	\myeq
	V_{\text{tree}}&=\left(\left\lvert\mu\right\rvert^{2}+m_{H_{u}}^{2}\right)\left\lvert H_{u}^{0}\right\rvert^{2}+\left(\left\lvert\mu\right\rvert^{2}+m_{H_{d}}^{2}\right)\left\lvert H_{d}^{0}\right\rvert^{2}-\left(bH_{u}^{0}H_{d}^{0}+\text{c.c.}\right)+\frac{\left(g^{2}+g'^{2}\right)}{8}\left(\left\lvert H_{u}^{0}\right\rvert^{2}-\left\lvert H_{d}^{0}\right\rvert^{2}\right)^{2}
	\myeqend
	where $b\equiv B\mu$.
	
	In the $\overline{\text{DR}}'$ scheme, to remove the dependence of the 1-loop effective potential on $m_{\epsilon}^{2}$, we can write\cite{Martin:2001vx}
	\myeq
	\Delta V&=\frac{1}{16\pi^{2}}V^{(1)}_{\overline{\text{DR}}'}
	\myeqend
	where
	\[
	V^{(1)}_{\overline{\text{DR}}'}=\sum\limits_{n}(-1)^{2s_{n}}\left(2s_{n}+1\right)h\left(m_{n}^{2}\right)\equiv\text{STr}(h(m_{n}^{2})),
	\]
	\[
	h(x)=\frac{x^{2}}{4}\left[\overline{\ln}(x)-\frac{3}{2}\right]
	\]
	and
	\[
	\overline{\ln}(x)\equiv\ln\left(\frac{x}{Q^{2}}\right)
	\]
	with $Q$ being the renormalization scale. One must also be careful to account for color multiplicity and charge multiplicity factors, where
	\[
	n_{\widetilde{f}}=\begin{cases}
		1,&\widetilde{f}=\text{color neutral sparticle}\\
		3,&\widetilde{f}=\text{squark}\\\end{cases}
	\]
	represents the number of colors of $\widetilde{f}$ and
	\[
	n_{\text{charge}}=\begin{cases}
		1, &\widetilde{f}\text{ is uncharged}\\
		2, &\widetilde{f}\text{ is charged}\\\end{cases}.
	\]
	
	For notation purposes, we may write
	\[
	v_{u}=\left\langle H_{u}^{0}\right\rangle\ \ \ {\rm and}\ \ \ 
	v_{d}=\left\langle H_{d}^{0}\right\rangle
	\]
	such that
	\[
	v_{u}^{2}+v_{d}^{2}=v^{2}=\frac{2m_{Z}^{2}}{g^{2}+g'^{2}}\approx(174\text{ GeV})^{2}.
	\]
	Then, defining the parameter $\beta$ through
	\[
	\tan(\beta)\equiv\frac{v_{u}}{v_{d}}
	\]
	one may write
	\[
	v_{u}=v\sin(\beta)\ \ \ {\rm and}\ \ \ v_{d}=v\cos(\beta).
	\]
	Therefore, the minimization conditions are:
	\myeq
	m_{H_{u}}^{2}+\Sigma_{u}^{u}+\left\lvert\mu\right\rvert^{2}-b\cot(\beta)-\frac{m_{Z}^{2}}{2}\cos(2\beta)&=0
	\myeqend
	and
	\myeq
	m_{H_{d}}^{2}+\Sigma_{d}^{d}+\left\lvert\mu\right\rvert^{2}-b\tan(\beta)+\frac{m_{Z}^{2}}{2}\cos(2\beta)&=0
	\myeqend
	where we treat $v_{u,d}$ as real variables in the differentiation such that the derivatives of $\Delta V(v_{u},v_{d})$ can be written as
	\myeq
	\Sigma_{u}^{u}&=\frac{\p\Delta V\left(v_{u},v_{d}\right)}{\p v_{u}^{2}}\\
	&=\frac{\p\Delta V}{\p v_{u}}\underbrace{\frac{\p v_{u}}{\p v_{u}^{2}}}_{\frac{1}{2v_{u}}}+\frac{\p\Delta V}{\p v_{d}}\underbrace{\frac{\p v_{d}}{\p v_{u}^{2}}}_{0}\\
	&=\frac{1}{2v_{u}}\frac{\p\Delta V}{\p v_{u}},
	\myeqend
	\myeq
	\Sigma_{d}^{d}&=\frac{\p\Delta V\left(v_{u},v_{d}\right)}{\p v_{d}^{2}}\\
	&=\frac{\p\Delta V}{\p v_{u}}\underbrace{\frac{\p v_{u}}{\p v_{d}^{2}}}_{0}+\frac{\p\Delta V}{\p v_{d}}\underbrace{\frac{\p v_{d}}{\p v_{u}^{2}}}_{\frac{1}{2v_{d}}}\\
	&=\frac{1}{2v_{d}}\frac{\p\Delta V}{\p v_{d}},
	\myeqend
	\myeq
	\Sigma_{u}^{d}&=\frac{\p\left(\Delta V\right)}{\p(v_{u}v_{d}+c.c.)},
	\myeqend
	with
	\myeq
	\Sigma_{u}^{d}&=\Sigma_{d}^{u}.
	\myeqend
	With these notes in mind, we can write the minimization conditions as
	\myeq
	m_{Z}^{2}&=\frac{\left\lvert m_{H_{d}}^{2}+\Sigma_{d}^{d}-m_{H_{u}}^{2}-\Sigma_{u}^{u}\right\rvert}{\underbrace{\sqrt{1-\sin^{2}(2\beta)}}_{\cos(2\beta)}}-m_{H_{u}}^{2}-\Sigma_{u}^{u}-m_{H_{d}}^{2}-\Sigma_{d}^{d}-2\left\lvert\mu\right\rvert^{2}
	\myeqend
	and
	\myeq
	\sin(2\beta)&=\frac{2b}{m^{2}_{H_{u}}+\Sigma_{u}^{u}+m_{H_{d}}^{2}+\Sigma_{d}^{d}+2\left\lvert\mu\right\rvert^{2}}
	\myeqend
	or equivalently
	\[
	b=\left[\left(m_{H_{u}}^{2}+\Sigma_{u}^{u}+\left\lvert\mu\right\rvert^{2}\right)+\left(m_{H_{d}}^{2}+\Sigma_{d}^{d}+\left\lvert\mu\right\rvert^{2}\right)\right]\sin(\beta)\cos(\beta).
	\]
	The $\Sigma$ contributions are, explicitly,
	\myeq
	\Sigma_{u}^{u}&=\sum\limits_{n}\frac{(-1)^{2s_{n}}}{32\pi^{2}}\left(2s_{n}+1\right)m_{n}^{2}\left(\frac{\p m_{n}^{2}}{\p v_{u}^{2}}\right)\left(\overline{\ln}\left(m_{n}^{2}\right)-1\right),\\
	\Sigma_{d}^{d}&=\sum\limits_{n}\frac{(-1)^{2s_{n}}}{32\pi^{2}}\left(2s_{n}+1\right)m_{n}^{2}\left(\frac{\p m_{n}^{2}}{\p v_{d}^{2}}\right)\left(\overline{\ln}\left(m_{n}^{2}\right)-1\right),\\
	\Sigma_{u}^{d}&=\Sigma_{d}^{u}\\
	&=\sum\limits_{n}\frac{(-1)^{2s_{n}}}{64\pi^{2}}\left(2s_{n}+1\right)m_{n}^{2}\left(\frac{\p m_{n}^{2}}{\p \left(v_{u}v_{d}\right)}\right)\left(\overline{\ln}\left(m_{n}^{2}\right)-1\right),
	\myeqend
	with
	\myeq
	F(m^{2})&=m^{2}\left[\overline{\ln}\left(m^{2}\right)-1\right].
	\myeqend
	The color and charge factors are accounted for as mentioned above, and each individual contribution is given below. 
	\subsection*{Squarks and sleptons}
	The stop squark squared mass matrix is given by
	\myeq
	\textbf{m}_{\widetilde{t}}^{2}&=\begin{bmatrix}
		m^{2}_{\widetilde{t}_{L}}+m_{t}^{2}+\Delta_{\widetilde{u}_{L}}&a^{*}_{t}v_{u}-\mu y_{t}v_{d}\\
		a_{t}v_{u}-\mu^{*}y_{t}v_{d}&m^{2}_{\widetilde{t}_{R}}+m_{t}^{2}+\Delta_{\widetilde{u}_{R}}\\\end{bmatrix}
	\myeqend
	where
	\myeq
	\Delta_{\phi}&=\left(T_{3\phi}-Q_{\phi}\sin^{2}(\theta_{W})\right)\left(\frac{g^{2}+g'^{2}}{2}\right)(v_{d}^{2}-v_{u}^{2})
	\myeqend
	where $T_{3\phi}$ is the third component of the weak isospin of $\phi$, and $Q_{\phi}$ is the electrical charge of $\phi$. Hence,
	\[
	\frac{\p\Delta_{\phi}}{\p v_{u}^{2}}=-(T_{3\phi}-Q_{\phi}\underbrace{\sin^{2}(\theta_{W})}_{x_{W}})\left(\frac{g^{2}+g'^{2}}{2}\right)
	\]
	and
	\[
	\frac{\p\Delta_{\phi}}{\p v_{d}^{2}}=\left(T_{3\phi}-Q_{\phi}x_{W}\right)\left(\frac{g^{2}+g'^{2}}{2}\right).
	\]
	At tree-level, the following mass relations hold for the running masses:
	\myeq
	m_{t}&=y_{t}v_{u},
	\myeqend
	\myeq
	m_{b}&=y_{b}v_{d},
	\myeqend
	\myeq
	m_{\tau}&=y_{\tau}v_{d}.
	\myeqend
	The eigenvalues of Eq. (A.14) are
	\myeq
	m^{2}_{\widetilde{t}_{1,2}}&=\frac{1}{2}\Bigg(m^{2}_{\widetilde{t}_{L}}+2m_{t}^{2}+m^{2}_{\widetilde{t}_{R}}+\Delta_{\widetilde{u}_{L}}+\Delta_{\widetilde{u}_{R}}\\
&\mp\sqrt{(m^{2}_{\widetilde{t}_{L}}-m^{2}_{\widetilde{t}_{R}}+\Delta_{\widetilde{u}_{L}}-\Delta_{\widetilde{u}_{R}})^{2}+4\left[|a_{t}|^{2}v_{u}^{2}-v_{d}v_{u}y_{t}\mu^{*}a_{t}^{*}+v_{d}v_{u}y_{t}\mu a_{t}+v_{d}v_{u}+|\mu|^{2}v_{d}^{2}y_{t}^{2}\right]}\Bigg)
	\myeqend
	where $m^{2}_{\widetilde{t}_{1}}<m^{2}_{\widetilde{t}_{2}}$.
	
	Note that
	\[
	g_{Z}^{2}\equiv\frac{g^{2}+\left(g'\right)^{2}}{8}\ \ {\rm so\ that}\ \ \ 
        m_{Z}^{2}\equiv 4 v^{2}\cdot g_Z^2 .
	\]
	The trilinear couplings are written in ``reduced" form, i.e.,
	\[
	a_{i}\equiv A_{i}\cdot y_{i}
	\]
	with $y_{i}$ the corresponding Yukawa coupling.
	
	The radiative correction terms are then
	\myeq
	\Sigma_{u}^{u}(\widetilde{t}_{1,2})&=\frac{3}{16\pi^{2}}F(m_{\widetilde{t}_{1,2}}^{2})\cdot\left[y_{t}^{2}-g_{Z}^{2}\mp\frac{a_{t}^{2}-2g_{Z}^{2}\cdot\Delta_{\widetilde{t}}}{m_{\widetilde{t}_{2}}^{2}-m_{\widetilde{t}_{1}}^{2}}\right]
	\myeqend
	with
	\[
	\Delta_{\widetilde{t}}\equiv2\left(\frac{1}{2}-\frac{4}{3}\sin^{2}(\theta_{W})\right)\cdot\left[\frac{m_{\widetilde{t}_{L}}^{2}-m_{\widetilde{t}_{R}}^{2}}{2}+\left(m_{Z}^{2}\cos(2\beta)\cdot\left[\frac{1}{4}-\frac{2}{3}\sin^{2}(\theta_{W})\right]\right)\right]
	\]
	and
	\myeq
	\Sigma_{d}^{d}(\widetilde{t}_{1,2})&=\frac{3}{16\pi^{2}}F(m_{\widetilde{t}_{1,2}^{2}})\left[g_{Z}^{2}\mp\frac{y_{t}^{2}\cdot\mu^{2}+2g_{Z}^{2}\cdot\Delta_{\widetilde{t}}}{m_{\widetilde{t}_{2}}^{2}-m_{\widetilde{t}_{1}}^{2}}\right]
	\myeqend
	where the minus (plus) sign corresponds to $\widetilde{t}_{1(2)}$.
	
	Next, with the bottom squark mass matrix
	\myeq
	\textbf{m}_{\widetilde{b}}^{2}&=\begin{bmatrix}
		m^{2}_{\widetilde{b}_{L}}+\Delta_{\widetilde{d}_{L}}&a^{*}_{b}v_{d}-\mu y_{b}v_{u}\\
		a_{b}v_{d}-\mu^{*}y_{b}v_{d}&m^{2}_{\widetilde{b}_{R}}+\Delta_{\widetilde{d}_{R}}\\\end{bmatrix},
	\myeqend
	the eigenvalues are computed to be
	\myeq
	\resizebox{0.9\hsize}{!}{$m^{2}_{\widetilde{b}_{1,2}}=\frac{1}{2}\left(m^{2}_{\widetilde{b}_{L}}+\Delta_{\widetilde{d}_{L}}+m^{2}_{\widetilde{b}_{R}}+\Delta_{\widetilde{d}_{R}}\mp\sqrt{(-m^{2}_{\widetilde{b}_{L}}-\Delta_{\widetilde{d}_{L}}+m^{2}_{\widetilde{b}_{R}}+\Delta_{\widetilde{d}_{R}})^{2}+4(a_{b}^{*}v_{d}-\mu v_{u}y_{t})(a_{b}v_{d}-\mu^{*}v_{u}y_{t})}\right).$}
	\myeqend
	Thus, the radiative correction terms are
	\myeq
	\Sigma_{u}^{u}(\widetilde{b}_{1,2})&=\frac{3}{16\pi^{2}}F(m_{\widetilde{b}_{1,2}}^{2})\cdot\left[y_{b}^{2}-g_{Z}^{2}\mp\frac{a_{b}^{2}-2g_{Z}^{2}\cdot\Delta_{\widetilde{b}}}{m_{\widetilde{b}_{2}}^{2}-m_{\widetilde{b}_{1}}^{2}}\right]
	\myeqend
	with
	\[
	\Delta_{\widetilde{b}}=2\left(\frac{1}{2}-\frac{2}{3}\sin^{2}(\theta_{W})\right)\cdot\left(\frac{m_{b_{L}}^{2}-m_{b_{R}}^{2}}{2}-m_{Z}^{2}\cos(2\beta)\cdot\left[\frac{1}{4}-\frac{1}{3}\sin^{2}(\theta_{W})\right]\right)
	\]
	and
	\myeq
	\Sigma_{d}^{d}(\widetilde{b}_{1,2})&=\frac{3}{16\pi^{2}}F(m_{\widetilde{b}_{1,2}}^{2})\cdot\left[g_{Z}^{2}\mp\frac{y_{b}^{2}\cdot\mu^{2}+2g_{Z}^{2}\cdot\Delta_{\widetilde{b}}}{m_{\widetilde{b}_{2}}^{2}-m_{\widetilde{b}_{1}}^{2}}\right].
	\myeqend
	Similarly, one can obtain the result for the staus by exchanging $b\to\tau$, $n_{\widetilde{\tau},\text{ col.}}\to1$, and
	\[
	\Delta_{\widetilde{\tau}}\equiv2\left(\frac{1}{2}-2\sin^{2}(\theta_{W})\right)\cdot\left[\frac{m_{\widetilde{\tau}_{L}}^{2}-m_{\widetilde{\tau}_{R}}^{2}}{2}-\left(\frac{1}{4}-\sin^{2}(\theta_{W})\right)\right].
	\]
	\subsection*{Sfermions}
	For a general sfermion $\widetilde{f}_{L,R}$ in the first or second generation, the masses can be parameterized based on the boundary conditions of the model. It can be shown that for these sfermions,
	\myeq
	\frac{\p m_{\widetilde{f}}^{2}}{\p v_{u,d}^{2}}=\overbrace{\underbrace{\pm}_{v_{u}^{2}}}^{v_{d}^{2}}(T_{3\widetilde{f}}-Q_{\widetilde{f}}x_{W})\left(\frac{g^{2}+g'^{2}}{2}\right)
	\myeqend
	which leads to the cancellation between first and second generation sfermion corrections, where the cancellation arises between $\Sigma_{u}^{u}(\widetilde{f})$ and $\Sigma_{d}^{d}(\widetilde{f})$. Indeed, one finds that for the squarks,
	\myeq
	\Sigma^{u}_{u}(\widetilde{f}_{L,R})&=\frac{-3}{4\pi^{2}}(T_{3\widetilde{f}_{L,R}}-Q_{\widetilde{f}_{L,R}}x_{W})g_{Z}^{2}F(m_{\widetilde{f}_{L,R}}^{2})
	\myeqend
	and
	\myeq
	\Sigma^{d}_{d}(\widetilde{f}_{L,R})&=\frac{+3}{4\pi^{2}}(T_{3\widetilde{f}_{L,R}}-Q_{\widetilde{f}_{L,R}}x_{W})g_{Z}^{2}F(m_{\widetilde{f}_{L,R}}^{2}).
	\myeqend
	For the sleptons in the 1st and 2nd generations, replace the color factor of $3$ in the numerator with $1$, and for the slepton sneutrinos, replace the color factor of $3\to1$ and the charge factor from $2\to1$. 
	\subsection*{Neutralinos}
	For the neutralinos, the \emph{unsquared} mass matrix can be given by the following, where $c_{\beta}=\cos(\beta),s_{\beta}=\sin(\beta)$, and likewise for other angles:
	\myeq
	\textbf{m}_{\widetilde{N}}&=\begin{bmatrix}
		M_{1}&0&-c_{\beta}s_{W}m_{Z}&s_{\beta}s_{W}m_{Z}\\
		0&M_{2}&c_{\beta}c_{W}m_{Z}&-s_{\beta}c_{W}m_{Z}\\
		-c_{\beta}s_{W}m_{Z}&c_{\beta}c_{W}m_{Z}&0&-\mu\\
		s_{\beta}s_{W}m_{Z}&-s_{\beta}c_{W}m_{Z}&-\mu&0\\
	\end{bmatrix}.
	\myeqend
	One could square this matrix and solve for the eigenvalues by brute force using the Ferrari method, and then differentiating the resultant squared mass eigenvalues. Instead, we use the method proposed by Ibrahim and Nath for taking the derivatives of eigenvalues\cite{Ibrahim:2002zk}. Consider the characteristic polynomial of the squared neutralino mass matrix. The coefficients will each be functions of $v_{u},v_{d}$. As such, one may write in general form
	\myeq
	F(\lambda)&=\lambda^{4}+b_{\lambda}(v_{u},v_{d})\lambda^{3}+c_{\lambda}(v_{u},v_{d})\lambda^{2}+d_{\lambda}(v_{u},v_{d})\lambda+e_{\lambda}(v_{u},v_{d})\\
	&=0.
	\myeqend
	Then the eigenvalues have derivatives given by
	\myeq
	\frac{\p m^{2}_{\widetilde{N}_{i}}}{\p v_{u,d}^{2}}&=\frac{-D_{v_{u,d}^{2}}F}{D_{\lambda} F}\Bigg\rvert_{\lambda=m^{2}_{\widetilde{N}_{i}}},
	\myeqend
	which can be obtained by taking the derivative of Eq. (A.30), with $i=1,2,3,4$,
	\myeq
	D_{\lambda}F&\equiv\frac{dF}{d\lambda},
	\myeqend
	and
	\myeq
	D_{v_{u,d}^{2}}F&=\frac{\p b_{\lambda}}{\p v_{u,d}^{2}}\lambda^{3}+\frac{\p c_{\lambda}}{\p v_{u,d}^{2}}\lambda^{2}+\frac{\p d_{\lambda}}{\p v_{u,d}^{2}}\lambda+\frac{\p e_{\lambda}}{\p v_{u,d}^{2}}/
	\myeqend
	From there, the explicit form of $\Sigma^{u,d}_{u,d}(\widetilde{N}_{i})$ for $i=1,2,3,4$ can be written, which will take the form
	\myeq
	\Sigma^{u,d}_{u,d}(\widetilde{N}_{i})&=\frac{-1}{16\pi^{2}}\left[\frac{-D_{v_{u,d}^{2}}F}{D_{\lambda} F}\Bigg\rvert_{\lambda=m^{2}_{\widetilde{N}_{i}}}\right]F(m^{2}_{\widetilde{N}_{i}})/
	\myeqend
	\subsection*{Charginos}
	The chargino mass matrix is in $2\times2$ block form on the off-diagonals, and can be squared to obtain the squared mass matrix,
	\myeq
	\textbf{m}_{\widetilde{C}}^{2}&=\begin{bmatrix}
		M_{2}^{2}+g^{2}v_{d}^{2}&g(M_{2}v_{u}+v_{d}\mu)&0&0\\
		g(M_{2}v_{u}+\mu v_{d})&g^{2}v_{u}^{2}+\mu^{2}&0&0\\
		0&0&M_{2}^{2}+g^{2}v_{u}^{2}&g(M_{2}v_{d}+\mu v_{u})\\
		0&0&g(M_{2}v_{d}+\mu v_{u})&g^{2}v_{d}^{2}+\mu^{2}\\
	\end{bmatrix} .
	\myeqend
	This has doubly degenerate eigenvalues
	\myeq
	m^{2}_{\widetilde{C}_{1,2}}&=\frac{1}{2}\left[|M_{2}|^{2}+|\mu|^{2}+g^{2}(v_{u}^{2}+v_{d}^{2})\mp\sqrt{\left[g^{2}(v_{u}+v_{d})^{2}+(M_{2}-\mu)^{2}\right]\left[g^{2}(v_{d}-v_{u})^{2}+(M_{2}+\mu)^{2}\right]}\right]
	\myeqend
	with $m^{2}_{\widetilde{C}_{1}}<m^{2}_{\widetilde{C}_{2}}$, leading to
	\myeq
	\Sigma_{u}^{u}(\widetilde{C}_{1,2})&=\frac{-g^{2}}{16\pi^{2}}\left(1\mp\frac{(-2)\cdot m_{W}^{2}\cdot\cos(2\beta)+M_{2}^{2}+\mu^{2}}{m^{2}_{\widetilde{C}_{2}}-m^{2}_{\widetilde{C}_{1}}}\right)F(m^{2}_{\widetilde{C}_{1,2}})
	\myeqend
	and
	\myeq
	\Sigma_{d}^{d}(\widetilde{C}_{1,2})&=\frac{-g^{2}}{16\pi^{2}}\left(1\mp\frac{2\cdot m_{W}^{2}\cdot\cos(2\beta)+M_{2}^{2}+\mu^{2}}{m^{2}_{\widetilde{C}_{2}}-m^{2}_{\widetilde{C}_{1}}}\right)F(m^{2}_{\widetilde{C}_{1,2}}).
	\myeqend
	with $m_{W}^{2}$ given below.
	\subsection*{Weak bosons}
	Using the mass relations
	\myeq
	m_{W}^{2}&=\frac{g^{2}}{2}v^{2}\implies \
	\frac{\p m_{W}^{2}}{\p v_{u,d}^{2}}&=\frac{g^{2}}{2}
	\myeqend
	and
	\myeq
	m_{Z}^{2}&=\frac{(g^{2}+g'^{2})}{2}(v^{2})\implies\ 
	\frac{\p m_{Z}^{2}}{\p v_{u,d}^{2}}&=\frac{g^{2}+g'^{2}}{2}
	\myeqend
	such that the radiative corrections are
	\myeq
	\Sigma^{u}_{u}(W^{\pm})&=\Sigma^{d}_{d}(W^{\pm})\
	&=\frac{3g^{2}}{32\pi^{2}}F(m_{W}^{2})
	\myeqend
	and
	\myeq
	\Sigma^{u}_{u}(Z^{0})&=\Sigma^{d}_{d}(Z^{0})\
	&=\frac{3(g^{2}+g'^{2})}{64\pi^{2}}F(m_{Z}^{2}).
	\myeqend
	\subsection*{Higgs bosons}
	For the Higgs bosons, we have the mass relations (with $m^{2}_{h_{0}}<m^{2}_{H_{0}}$)
	\myeq
	m^{2}_{h_{0},H_{0}}&=\frac{1}{2}\left(m_{A_{0}}^{2}+m_{Z}^{2}\mp\sqrt{(m_{A_{0}}^{2}-m_{Z}^{2})^{2}+4m_{Z}^{2}m_{A_{0}}^{2}\sin^{2}(2\beta)}\right)
	\myeqend
	and
	\myeq
	m^{2}_{A_{0}}&=\frac{2b}{\sin(2\beta)}\
	&=2|\mu|^{2}+m^{2}_{H_{u}}+m^{2}_{H_{d}}
	\myeqend
	Therefore,
	\myeq
	\Sigma^{u}_{u}(h_{0},H_{0})&=\frac{g_{Z}^{2}}{16\pi^{2}}\left[1\mp\frac{m_{Z}^{2}+m_{A_{0}}^{2}\cdot\left(1+4\cos(2\beta)+2\cos^{2}(2\beta)\right)}{m^{2}_{H_{0}}-m^{2}_{h_{0}}}\right]F(m_{h_{0},H_{0}}^{2})
	\myeqend
	and
	\myeq
	\Sigma^{d}_{d}(h_{0},H_{0})&=\frac{g_{Z}^{2}}{16\pi^{2}}\left[1\mp\frac{m_{Z}^{2}+m_{A_{0}}^{2}\cdot\left(1-4\cos(2\beta)+2\cos^{2}(2\beta)\right)}{m^{2}_{H_{0}}-m^{2}_{h_{0}}}\right]F(m_{h_{0},H_{0}}^{2}).
	\myeqend
	Then, for the charged Higgs, with the relation
	\myeq
	m^{2}_{H^{\pm}}&=m^{2}_{A_{0}}+m_{W}^{2}
	\myeqend
	one simply finds that 
	\myeq
	\Sigma^{u,d}_{u,d}(H^{\pm})&=\frac{g^{2}}{32\pi^{2}}F(m^{2}_{H^{\pm}}).
	\myeqend
	\subsection*{SM fermions}
	Finally, we can use the mass relations mentioned in Eqs. (16) - (18) to obtain the top, bottom, and $\tau$ contributions:
	\myeq
	\Sigma^{u}_{u}(t)&=\frac{-3y_{t}^{2}}{16\pi^{2}}F(m_{t}^{2}),
	\myeqend
	\myeq
	\Sigma^{d}_{d}(t)&=0,
	\myeqend
	\myeq
	\Sigma^{u}_{u}(b)&=0,
	\myeqend
	\myeq
	\Sigma^{d}_{d}(b)&=\frac{-3y_{b}^{2}}{16\pi^{2}}F(m_{b}^{2}),
	\myeqend
	\myeq
	\Sigma^{u}_{u}(\tau)&=0,
	\myeqend
	and
	\myeq
	\Sigma^{d}_{d}(\tau)&=\frac{-y_{\tau}^{2}}{16\pi^{2}}F(m_{\tau}^{2}).
	\myeqend
	We choose
	\myeq
	Q=\sqrt{m_{\text{stop}_1}m_{\text{stop}_{2}}}
	\myeqend
	as the renormalization scale used in the function $F$.
	
	From these terms, one can use the minimization conditions to define the naturalness measure $\Delta_{EW}$:
	\myeq
	\Delta_{EW}&\equiv\frac{2\max\left\lvert C_{i}\right\rvert}{m_{Z}^{2}}
	\myeqend
	where the $C_{i}$ terms are the individual contributions from the Higgs minimization condition ($i=H_{u},H_{d},\mu,\Sigma_{u}^{u},\Sigma_{d}^{d}$). Specifically, $C_{\mu}=\left\lvert\mu\right\rvert^{2}$,
	\[
	C_{H_{u}}=\frac{-m_{H_{u}}^{2}\tan^{2}(\beta)}{\tan^{2}(\beta)-1}\ \ {\rm and}\ \
	C_{H_{d}}=\frac{m_{H_{d}}^{2}}{\tan^{2}(\beta)-1},
	\]
and
	\[
	C_{\Sigma_{u}^{u}}=\frac{-\Sigma_{u}^{u}\tan^{2}(\beta)}{\tan^{2}(\beta)-1}\ \ {\rm with}\ \ 
	C_{\Sigma_{d}^{d}}=\frac{\Sigma_{d}^{d}}{\tan^{2}(\beta)-1}.
	\]


\bibliography{comp}
\bibliographystyle{elsarticle-num}

\end{document}